\documentclass[runningheads]{llncs}

\usepackage[mobile]{eccv}

\usepackage{eccvabbrv}

\usepackage{placeins} 
\usepackage{siunitx}

\usepackage{tikz}
\usepackage{pgfplots}
\pgfplotsset{compat=1.12}
\usepgfplotslibrary{fillbetween}
\usetikzlibrary{patterns}

\usepackage{graphicx}

\usepackage{tabularx,ragged2e,booktabs}
\newcolumntype{L}{>{\RaggedRight\arraybackslash}X} 

\usepackage{longtable}

\usepackage{amsmath}
\usepackage{adjustbox}

\usepackage{subcaption}
\usepackage{mathtools}
\usepackage{nicefrac}
\usepackage{multirow}

\usepackage{etoolbox}
\AtBeginEnvironment{bmatrix}{\setlength{\arraycolsep}{2pt}}

\usepackage{listings}
\usepackage{multicol}
\definecolor{codegreen}{rgb}{0,0.6,0}
\definecolor{codegray}{rgb}{0.5,0.5,0.5}
\definecolor{codepurple}{rgb}{0.58,0,0.82}
\definecolor{backcolour}{rgb}{0.95,0.95,0.92}
\lstdefinestyle{mystyle}{
    backgroundcolor=\color{backcolour},
    commentstyle=\color{codegreen},
    keywordstyle=\color{magenta},
    numberstyle=\tiny\color{codegray},
    stringstyle=\color{codepurple},
    basicstyle=\ttfamily\footnotesize,
    breakatwhitespace=false,
    breaklines=true,
    captionpos=b,
    keepspaces=true,
    numbers=left,
    numbersep=5pt,
    showspaces=false,
    showstringspaces=false,
    showtabs=false,
    tabsize=2
}
\usepackage{lscape}

\DeclareMathOperator{\EX}{\mathbb{E}}
\DeclareMathOperator{\prob}{\mathsf{P}}
\DeclareMathOperator{\snr}{\mathsf{SNR}}

\interfootnotelinepenalty=\@M

\usepackage[accsupp]{axessibility}  

\usepackage{hyperref}

\usepackage{orcidlink}

\begin{document}

\title{Photon Inhibition for Energy-Efficient Single-Photon Imaging}

\author{Lucas J. Koerner\inst{1}\orcidlink{0000-0002-7236-7202} \and
Shantanu Gupta\inst{2}\orcidlink{0009-0008-3772-4316}\and
Atul Ingle\inst{3}\orcidlink{0000-0002-3695-5891} \and
Mohit Gupta\inst{2}\orcidlink{https://orcid.org/0000-0002-2323-7700}}

\authorrunning{L.~Koerner et al.}

\institute{Department of Electrical and Computer Engineering, University of St. Thomas, St. Paul, MN, 55105, USA \email{koerner.lucas@stthomas.edu}	\and
Department of Computer Sciences, University of Wisconsin-Madison, Madison, WI, 53706, USA \email{\{sgupta,mohitg\}@cs.wisc.edu}	\and
Department of Computer Science, Portland State University, Portland, OR, 97201, USA \email{ingle2@pdx.edu}}

\maketitle

\begin{abstract}
Single-photon cameras (SPCs) are emerging as sensors of choice for various challenging imaging applications. One class of SPCs based on the single-photon avalanche diode (SPAD) detects individual photons using an avalanche process; the raw photon data can then be processed to extract scene information under extremely low light, high dynamic range, and rapid motion. Yet, single-photon sensitivity in SPADs comes at a cost --- each photon detection consumes more energy than that of a CMOS camera. This avalanche power significantly limits sensor resolution and could restrict widespread adoption of SPAD-based SPCs. We propose a computational-imaging approach called \emph{photon inhibition} to address this challenge. Photon inhibition strategically allocates detections in space and time based on downstream inference task goals and resource constraints. We develop lightweight, on-sensor computational inhibition policies that use past photon data to disable SPAD pixels in real-time, to select the most informative future photons. As case studies, we design policies tailored for image reconstruction and edge detection, and demonstrate, both via simulations and real SPC captured data, considerable reduction in photon detections (over 90\% of photons) while maintaining task performance metrics. Our work raises the question of ``which photons should be detected?'', and paves the way for future energy-efficient single-photon imaging.
Source code for our experiments is available at \url{https://wisionlab.com/project/inhibition}.

\end{abstract}

\section{Introduction}
\begin{figure}[t!]
	\centering
	\includegraphics[width=\textwidth]{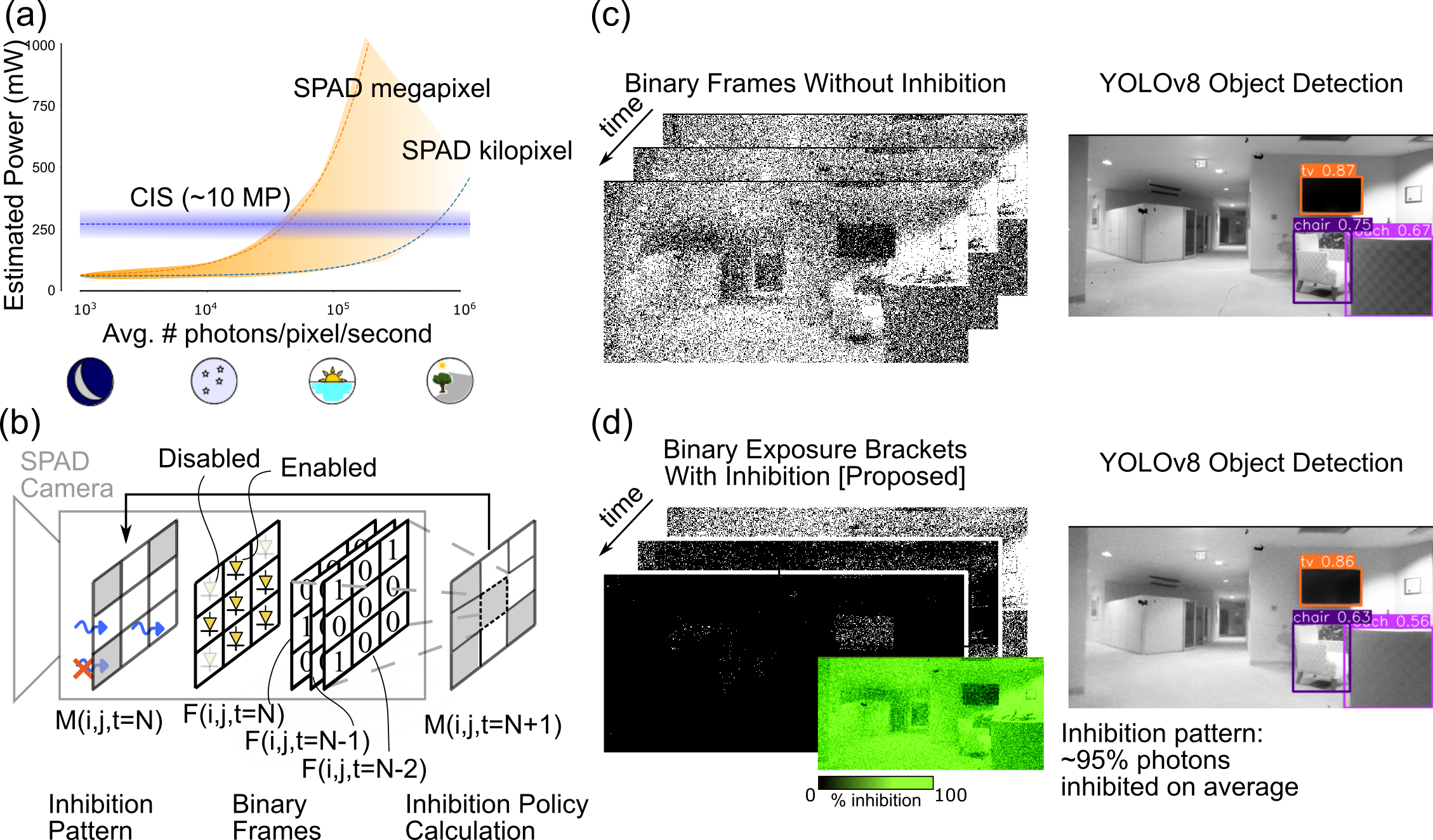}
	\caption{{\bf Photon inhibition for resource-efficient passive SPAD imaging.}
	(a) Unlike conventional CMOS cameras (CIS), the energy consumption in SPAD cameras increases with scene brightness, severely limiting the applicability of high resolution SPAD cameras in resource-constrained applications.
	(b) We expand the conventional imaging pipeline to incorporate ``inhibition'' that electronically enables or disables individual pixels to limit bandwidth and power consumption.
	Our method relies on lightweight mathematical operations called ``inhibition policies'' that update the inhibition patterns based on the history of photon detections.
	Inhibition policies can be optimized for image SNR or for downstream vision tasks.(c,d) Object detection, a high-level vision task, is successful even with a large fraction of photons inhibited.
	}
\label{fig:overview}
\end{figure}

From vacuum tube-based single-photon detectors invented in the early 20$\text{th}$ century \cite{iams1935secondary}, to 3D depth sensing LiDAR cameras found in today's smartphones \cite{yoshida2021}, single-photon camera (SPC) technology has come a long way in terms of pixel resolution and commercial availability for a variety of applications.
Thanks to CMOS-compatible single-photon avalanche diode (SPAD) arrays, SPCs are being increasingly used not only in niche fields such as scientific imaging and biomedical microscopy \cite{ulku512512SPAD2019}, but more widely for consumer photography \cite{morimotoMegapixel3DStackedCharge2021}.
SPAD-based SPCs have recently been fabricated into kilo-to-megapixel format arrays that are now commercially available \cite{morimotoMegapixel3DStackedCharge2021,yoshida2021}.
The extreme sensitivity and high speed can benefit passive low-light computer vision tasks, particularly in the presence of rapid scene or camera motion \cite{maBurstVisionUsing2023}, enable wide dynamic-range imaging \cite{duttonHighDynamicRange2018, liuSinglePhotonCameraGuided2022} and photon-starved active imaging applications such as 3D imaging (LiDAR) \cite{gyongy2021direct} and fluorescence microscopy \cite{ulku512512SPAD2019}.

SPAD camera pixels detect individual photons with extremely high frame rates by exploiting avalanche multiplication.
On one hand, being able to detect individual photons opens up new possibilities and capabilities for computer vision systems.
On the other hand, this also presents a unique challenge: Every photon-induced avalanche comes with a non-negligible energy cost, which is a challenge exclusive to SPAD-based cameras.
This flux-dependent photon detection power is a significant fraction of total power consumption in today's SPAD cameras and impedes further increases in their spatial resolution \cite{henderson25625640nm2019, morimotoMegapixel3DStackedCharge2021, ogi250fps124dBDynamicRange2021, ota37W143dBDynamicRange1Mpixel2022, takatsuka36UmpitchSPAD2023}.
For example, extrapolating the avalanche power of a recent SPAD sensor~\cite{takatsuka36UmpitchSPAD2023} to $\sim$10's of megapixels format predicts a power consumption of several watts in bright light\footnote{It has been shown, perhaps counter-intuitively, that SPADs do not saturate even under extremely bright conditions \cite{inglePassiveInterPhotonImaging2021b, ingleHighFluxPassive2019}.
Therefore, SPADs are not restricted to low-flux environments, but are being considered for vision applications across a wide dynamic range of lighting conditions (e.g., from a dark tunnel to bright sunlight).}
(as illustrated in \cref{fig:overview}(a)), far exceeding the power consumption of modern CMOS image sensors (CIS) of around \SI{300}{mW} \cite{shenzhencmtechnologycompanyltd16MPMIPI2023}.

We address this problem by being selective about which photons are detected on the SPAD sensor while maintaining good performance on various computer vision tasks.
To reduce avalanche power consumption and decouple it from photon flux, we propose a computational imaging technique called \emph{photon inhibition} where individual SPAD pixels are disabled adaptively based on previous photon detections in their spatio-temporal neighborhoods.
Electrically disabling pixels prevents photon detections to inhibit any avalanche power or subsequent processing. We implement lightweight on-sensor computations called \emph{inhibition policies} (\cref{fig:overview}(b)) to determine, in real-time and at single-pixel and single-frame granularity, which SPAD pixels to enable or disable.
Our method is inspired by retinal pre-processing of the human visual system where retinal neurons aggregate photon information over small spatio-temporal neighborhoods to cause neighboring retinal cells to become less sensitive to incident light \cite{doi:10.1146/annurev-vision-102016-061345, frankeInhibitionDecorrelatesVisual2017}.\footnote{We borrow the term ``inhibition'' from the phenomenon of ``lateral inhibition'' found in biological vision systems \cite{barlow1953summation}.}

Given that SPADs introduce a new challenge of flux-dependent power consumption, we establish, from first principles, novel \emph{energy-aware} imaging performance metrics for \textit{resource-constrained} single-photon imaging.
Based on these metrics, we design families of inhibition policies that distribute photon detections in space and time based on imaging / vision task goals and energy consumption constraints.
A critical consideration in the design of inhibition policies stems from the observation that these policies are meant to control (enable / disable) the sensor (\cref{fig:overview}(b)).
Therefore, it is \emph{essential} for these policies to be extremely lightweight since these need to be implemented on sensor with very limited compute and memory resources.
Furthermore, inhibition policies must execute at ultra-low latency to keep up with high-speed photon detections (reaching up to 100~kHz.).
Fortunately, since the raw data output from a SPAD-based SPC consists of binary-valued image frames, SPCs are naturally suited to real-time calculations at the image sensor plane under tight compute and memory budgets.
The proposed inhibition policies are lightweight, requiring only simple arithmetic and Boolean operations computed over local spatio-temporal neighborhoods, and thus amenable to in-pixel implementation \cite{carey100000Fps2013a, ardeleanComputationalImagingSPAD}.

In simulations and real experimental data, we show that our inhibition policies allocate photon detections to sensor pixels in a way that reduces detection energy for a given accuracy level of various vision tasks.
Our results show consequential energy savings when compared to a conventional capture scheme for tasks of
(i) image reconstruction: 42\% fewer photon detections at equal image quality;
(ii) edge detection: an edge sensitive inhibition policy reduces detections by 30\% at equal F-score; and
(iii) YOLOv8 object detection: remains successful with 95\% of photons inhibited under camera motion of a real-world SPC (\cref{fig:overview}(c,d)).
Through experiments with photon streams captured using real SPAD camera hardware over a wide range of illumination conditions, we show that our proposed inhibition policies preserve low-light details and, in bright-light, decouple flux and detection energy.

\smallskip
\noindent
\textbf{Scope and Limitations:}
There are several competing image sensor technologies today that resolve single-photons while capturing binary-valued raw frames at rates exceeding thousands of frames/second.
SPCs based on ``jots'' \cite{maPhotonnumberresolvingMegapixelImage2017} that do not rely on avalanche multiplication do not suffer from flux-dependent power consumption as SPADs.
Jots are an exciting technology, especially in scenarios that require high resolution and high dynamic range imaging \cite{ma19eRmsRead2021}.
In this paper, we focus on SPADs, due to their rapidly rising availability and commercial interest \cite{yoshida2021,canonspad}, and benefits over conventional CMOS sensors, both in low-light and bright scenes for a variety of computer vision tasks \cite{chi2020dynamic,maBurstVisionUsing2023}.

Our goals in this paper are to
(a) raise the question of ``which photons should be detected?'' in the context of energy-efficient single-photon imaging,
(b) establish a design space and metrics to evaluate various inhibition policies, and
(c) propose \emph{plausible} policies that respect practical hardware limitations for future on-chip implementation.
We emphasize that the inhibition policies proposed in this paper are not necessarily optimal.
This work is just a first step towards demonstrating that it is possible to achieve high performance with SPADs, while maintaining low power consumption via photon inhibition.

\section{Related Work \label{sec:related}}
\noindent{\bf Hardware approaches for reduced energy consumption:}
There is a strong dependence of SPAD power consumption on the pixel size --- the smaller the pixel, the lower the avalanche energy \cite{morimoto2021scaling}.
Although recent developments in SPAD pixel technology have reduced pixel sizes to below \SI{4}{\um} \cite{ogi2023iisw}, avalanche energy still contributes a large fraction of the total power consumption in a SPAD sensor \cite{severini2021spad}.
SPAD design optimizations have reduced the charge per avalanche by RF modulation of the bias voltage \cite{wayne2021low}, minimization of the junction capacitance \cite{ota37W143dBDynamicRange1Mpixel2022}, and smart \cite{boso2015low} and fast \cite{xu2020compact} quenching circuits.
Circuit architectures may require spatial and/or temporal co-incidence \cite{gyongy2021high} to reduce energy downstream in the processing chain, but avalanche energy remains.
Our work complements existing hardware approaches by preventing avalanches altogether to reduce illumination-dependent energy consumption.

\noindent{\bf SPAD dead-time and clocked recharge:}
SPADs require recharge after an avalanche-inducing photon detection during which subsequent photons are not recorded.
This dead-time inhibits photons at high exposure \cite{henderson25625640nm2019, ingleHighFluxPassive2019}, yet, power consumption remains excessive when the average inter-photon arrival interval is shorter than the SPAD dead time \cite{ota37W143dBDynamicRange1Mpixel2022, morimotoMegapixelTimegatedSPAD2020}.
Power consumption at high photon flux can be reduced by controlling the global rate of SPAD recharge (``clocked recharge'')  \cite{morimotoMegapixelTimegatedSPAD2020,takatsuka36UmpitchSPAD2023}.
Within a single recharge period a SPAD detects at most a single photon; subsequent photon arrivals do not induce an avalanche, thus power is reduced.
To further limit avalanches, clocked recharge has been combined with a limit on the number of detections and time to saturation circuitry to measure the intensity of the saturated pixels \cite{ogi250fps124dBDynamicRange2021}, coarse pixel-wise exposure control \cite{ota37W143dBDynamicRange1Mpixel2022}, and a sequence of recharge periods similar to exposure bracketing\cite{takatsuka36UmpitchSPAD2023}.
These methods could be considered special cases of inhibition policies that are global and do not adapt to scene content, and therefore, are limited in flexibility to trade power and measurement fidelity.
Consequently, in these methods, the power of avalanches remains a considerable fraction of total SPAD sensor power consumption \cite{takatsuka36UmpitchSPAD2023}.
For example, in this work we show considerable power savings, reaching up to $90\%$, for an object detection task.

\noindent{\bf Resource-aware imaging:}
Event-based vision sensors reduce power consumption by only transmitting scene information when an intensity change is detected \cite{gallego2020event}.
This idea has recently been applied to SPAD arrays to reduce power consumption due to data transfer \cite{della2020128,Sundar:2023:SoDaCam}.
In contrast, our method reduces power due to detection by selectively disabling photodiodes based on photon history over small spatial and temporal neighborhoods.
Miniaturized cameras with constraints on compute energy have transferred processing to the optical domain \cite{koppalWideAngleMicrovisionSensors2013}.
While we focus on passive imaging, depth sensing with SPADs and an active pulsed light source has related constraints such as acquisition time and laser power.
Adaptive gating reduces acquisition time \cite{po2022adaptive} and optimal allocation of the laser dwell time among pixels improves data quality \cite{medinBinomialNegativeBinomial2019, tilmonEnergyEfficientAdaptive3D2023a}.

\section{Observation Model \label{sec:imaging_background}}
During an exposure time $T$, a photon flux of $\phi$ results in an average number of photon conversions, or exposure, of $H = \phi T$ (we fold the sensor's \emph{photon detection probability}, or \emph{PDP}, into the definition of $\phi$, meaning it represents an effective flux).
The distribution of photon conversions, $K$, follows the Poisson distribution given by $\prob(K=k;H) = \frac{H^k e^{-H}}{k!}.$
During each binary exposure period, a SPAD pixel records a `1' if at least one photon was incident during that period, and `0' otherwise.
The probability of detecting at least one photon is given by $Y \coloneqq 1-\prob(K=0 ; H) = 1-e^{-H}$.
Hence, the SPAD pixel readout in each binary frame is a Bernoulli random variable $B \sim \mathsf{Bernoulli}(Y)$.
Multiple exposure time windows, or measurements ($W$), are recorded to reduce noise with the total number of detections
\begin{equation}
	D \coloneqq \left[\sum_{n=1}^{W} B_n\right] \sim \mathsf{Binomial}(W, 1-e^{-H}). \label{eq:num_det}
\end{equation}
We estimate the probability of detection and flux from a measurement $D$ as
\begin{align}
	\widehat{Y} 	&= 		\frac{D}{W} \label{eq:rate}\\
	\widehat{H} 	&= 		-\ln(1-\widehat{Y}). \label{eq:exp}
\end{align}

\smallskip
\noindent \textbf{Changes under inhibition:} Inhibition is represented by a binary state variable $M$ at every pixel, with $M_n = 1$ denoting enabled for the $n$-th measurement period.
$M$ is a random variable when an adaptive or data-dependent inhibition policy is used.
The total number of measurements changes to $W_{inh.} \coloneqq \sum_{n=1}^W M_n \le W$.
The number of detections becomes
\begin{equation}
	D_{inh.} \coloneqq \left[ \sum_{n=1}^W M_n B_n \right]
			= \left[\sum_{\substack{n=1\\M_n = 1}}^W B_n\right]
			\sim \mathsf{Binomial}(W_{inh.}, 1-e^{-H}),
	\label{eq:num_det_inhibited}
\end{equation}
and the flux is estimated similar to \cref{eq:rate,eq:exp}.
The second summation conveys that, with inhibition, the measurements when the pixel is enabled match the original model in \cref{eq:num_det} -- a result of the memoryless property of the Poisson arrival process. The model above requires that transitions in $M$ are synchronized with the clock signal used to gate the exposure, so that the PDP is not changed by inhibition. A second assumption is that the time for SPAD recharge is small relative to the clock period.
This is a desired property for passive SPAD-based imaging, and holds for many state-of-the-art SPADs \cite{takatsuka36UmpitchSPAD2023,severini2021spad,ota37W143dBDynamicRange1Mpixel2022}. It implies that PDP is approximately constant in time and does not depend on prior pixel state.

\section{Energy-Aware Performance Metrics \label{sec:efficiency_metrics}}
The exposure-referred signal-to-noise ratio ($\snr_H$) is commonly used to evaluate single-photon sensor performance, and can be computed as the ratio of the true exposure $H$ and the root-mean-squared error in the estimated exposure $\sqrt{\EX[(\widehat{H}-H)^2]}$ \cite{yang2011bits,fossumModelingPerformanceSingleBit2013}:
\begin{equation}
	\snr_H = \frac{H}{\sqrt{\EX[(\widehat{H}-H)^2]}}  = H \sqrt{\frac{W}{e^{H}-1}}.
\end{equation}
At low incident flux, $\snr_H$ is low due to shot noise.
$\snr_H$ improves as the likelihood of a photon detection increases until, in bright light with $H>1.6$, the $\snr_H$ degrades due to ``soft'' saturation of the response \cite{maQuantaBurstPhotography2020a, ingleHighFluxPassive2019}.

We propose two new \emph{energy-aware} modifications to $\snr_H$ to incorporate SPAD energy costs.
First, we propose a detection efficiency metric $\snr_{H/D}^2$ defined as the square of SNR normalized by the expected number of detections $\EX[D] = W(1-e^{-H})$:
\begin{equation}
	\snr_{H/D}^2 	:= \frac{\snr_H^2}{\EX[D]}
					= \frac{H^2 e^{-H}}{(1-e^{-H})^2}.
\end{equation}
\cref{fig:snr_and_tuning}(a) shows the $\snr_H$ (black) and the detection efficiency (red) versus the average photon arrivals per period ($H$).
When $H \ll 1$ all detections contribute significant information and $\snr_H^2$ increases linearly, similar to an ideal non-saturating sensor only limited by Poisson noise. Accordingly, $\snr_{H/D}^2 \approxeq 1$, the upper bound of this metric.
At larger exposure values, beginning around $H\approx 0.5$, the sensor begins to saturate which slows the growth of $\snr_H$ and thus degrades the detection efficiency.

A separate constraint is the total number of recharge periods during which a pixel is enabled and can measure either `0' or `1'.
This number of measurements ($W$) may be limited due to the energy to read out a frame, the depth of an in-pixel counter, and/or the maximum allowable sensing latency due to motion blur.
We establish a second metric, \textit{measurement efficiency} defined as the square of SNR normalized by the number of measurement windows:
\begin{equation}
	\snr_{H/W}^2 	:= 		\frac{\snr_H^2}{W}
					= 		\frac{H^2 e^{-H}}{1 - e^{-H}}. \label{eq:meas_eff}
\end{equation}
Fig.~\ref{fig:snr_and_tuning}(a) (blue) shows sub-optimal measurement efficiency at both low and high exposures with the best efficiency at $H=1.59$, $Y=0.80$, as demonstrated in \cite{chanWhatDoesOneBit2022}.

In an oracle setting with a known image, one can analytically derive non-uniform allocations of measurements $W_i$ across the pixels $i$, constraining the total expected detections to a fixed value and optimizing $\snr_H$ or mean-squared-error (see the supplement).
The loss metric may also be defined relative to the binary rate $Y$ instead of $H$ as above. In that case a useful base metric could be \emph{entropy} instead of $\snr_H$ \cite{gnanasambandamExposureReferredSignaltoNoiseRatio2022}.

\section{Spatio-temporal Inhibition Policies \label{sec:inhibition_policies}}
We now propose policies that calculate a spatio-temporal inhibition pattern for each pixel and each frame based on the history of photon frames and patterns.
Following \cref{sec:imaging_background}, we define an inhibition pattern using a binary-valued tensor $M$, where $M(i,j,t) = 0$ if pixel $(i,j)$ is disabled in the $t^\text{th}$ frame, and $M(i,j,t) = 1$ otherwise.
All pixels are initially enabled, and on-sensor calculations modify $M$ over time.
The binary photon cube is defined as $F(i,j,t)=1$ if the pixel is enabled (\ie, $M(i,j,t) = 1$) \emph{and} a photon is detected at pixel location $(i,j)$ in the $t^\text{th}$ frame, otherwise it is zero.

\renewcommand{\tabularxcolumn}[1]{>{\small}m{#1}}
\begin{figure}[!htbp]
\centering
	\begin{tabularx}{0.75\columnwidth}{ X }
		\adjustbox{max width=1\linewidth}{
	\begin{tikzpicture}[baseline=(current bounding box.center)]
		\newcommand\phht{0.75}
		\newcommand{\upph}[2]{\draw[->, thick] (#1, 0) -- (#1, #2)}
		\newcommand{\upphin}[2]{\draw[->, thin] (#1, 0) -- (#1, #2) node[midway]{X}}
		\newcommand{\expline}[3]{\draw[-, thin, color=white] (#1, 1.75) -- (#1, 2.25); \draw[<->] (#1, 2) -- (#1+#2, 2) node[midway,above] {#3}}

  		\node (L) at (0.25, 3.2) {};

		\node (A) at (-1, 1.5) {Inhibit};
        \node (Z) at (-1, 1.2) {Window};
		\node (B) at (-1, 0.5) {Photons};
		\node (O) at (0,0) {};
		\node (D) at (5, 0) {};
		\node (D2) at (10.9, -0.3) {time};
		
		\draw[->] (0,0) to (11,0);
		\expline{0}{1.5}{$T$};
		\expline{1.5}{1.5}{$T$};
		\expline{3}{1.5}{$T$};
		\expline{4.5}{1.5}{$T$};
		\expline{6}{1.5}{$T$};
		\expline{7.5}{1.5}{$T$};
		\expline{9}{1.5}{$T$};
		
		\upph{1.75}{\phht};
		\draw[draw=none, fill=black!15] (1.75,1) rectangle (3,1.75);
		
		\upph{3.75}{\phht};
		\draw[draw=none, fill=black!15] (3.75,1) rectangle (4.5,1.75);
		
		\node (L1) at (3.75, 2.75) {$S > \eta$};	
		\draw [->] (3.75,2.65) to [out=250,in=70] (4.5, 1.75);
		\draw[draw=none, fill=black!15] (4.5,1) rectangle (9,1.75);
		\draw[<->] (4.5, 2.75) -- (9, 2.75) node[midway,above] {$\tau_H$};
		
		\upphin{2.75}{\phht};
		\upphin{4.25}{\phht};
		
		\upphin{5.25}{\phht};
		\upphin{6.25}{\phht};
		\upphin{7}{\phht};
		\upphin{7.5}{\phht};
		
		\upph{10}{\phht};
		\draw[draw=none, fill=black!15] (10,1) rectangle (10.5,1.75);
		
	\end{tikzpicture}
 }
	\end{tabularx}
	\caption{\textbf{Calculation-based inhibition with dead time.}
		Arrows represent photons with an 'X' for inhibition.
		$T$ indicates the clocked recharge period.
		A score, $S$, calculated from past frames determines if future measurements are enabled or disabled.}
	\label{table:computation_policies}
\end{figure}
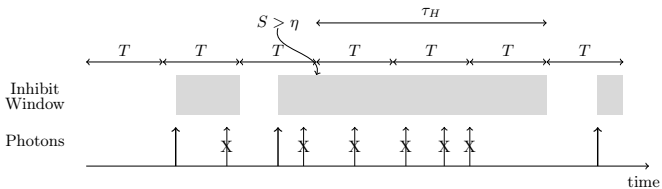

\begin{figure}[tb!]
\centering
	\includegraphics[width=\textwidth]{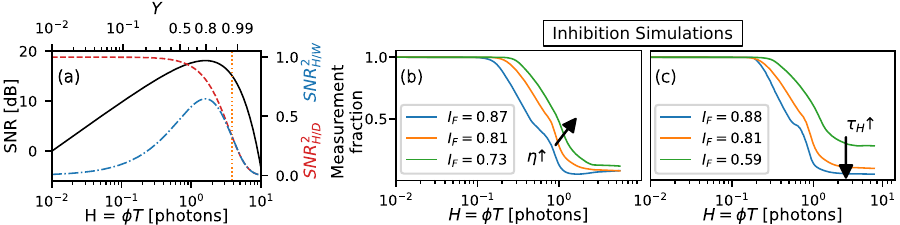}
	\caption{\textbf{Efficiency metrics and inhibition policies that track the metrics:}
	(a) The $\snr_H$ in dB (black), the detection efficiency (red, $--$), and the measurement efficiency (blue, {$- \cdot$}$-$) versus the exposure with $W=100$ measurements.
	The binary rate $Y = 1 - e^{-H}$ is indicated on the top axis.
	The vertical dotted line indicates the exposure and the binary rate at which the $\snr_H$ degrades by \SI{3}{dB} from the peak $\snr_H$.
	(b,c) Monte Carlo simulations of binary images using calculation-based inhibition policies demonstrate how the allocation of measurements versus the pixel exposure depends upon tuning parameters.
	(b) the inhibition threshold $\eta$ adjusts the exposure level at which pixels are inhibited to allow the measurement fraction (the ratio of active measurements to total number of frames) to follow the $\snr^2_{H/D}$ curve in (a).
	A smaller threshold more aggressively inhibits photons.
	(c) demonstrates the impact of the hold-off time, $\tau_H$, on the number of measurements allocated to the brightest pixels.
	The legend indicates the total fraction of photons inhibited as $I_F$. (b,c) show smoothed curves (Lowess filter, fraction of 1/5) of the measurement fraction vs. H.}
	\label{fig:snr_and_tuning}
\end{figure}

\cref{fig:overview}(b) shows the components of a photon inhibition processing layer, including the binary frames, $F$, and inhibition pattern $M$.
For ease of on-sensor implementation, we focus on policies that operate on small and local spatio-temporal neighborhoods of fixed sizes.
We rely on local arithmetic and Boolean computations and comparison operations, consistent with current in-pixel computational capabilities \cite{henderson25625640nm2019, charbon3DStackedCMOSSPAD2018, ardeleanComputationalImagingSPAD}.
\cref{table:computation_policies} shows a proposed on-sensor calculation approach that operates in a streaming fashion as frames accumulate to calculate an inhibition score, $S$, as the result of a spatio-temporal filter of the binary frames and inhibition pattern.
The score at each pixel is calculated as
\begin{equation}
	S(i,j,t) 	= 	K \ast [(2F(i,j,t)-\mathbf{1})\cdot M(i,j,t)] \label{eq:inhibition_score}
\end{equation}
which applies a spatio-temporal filtering kernel, $K$, of dimensions $L,H,T$ to a ternary representation of the pixel result ($1$, $0$, or $-1$ for a detection, a disabled pixel, or a measurement that does not detect a photon, respectively).
The kernel $K$ can typically be separated into spatial and temporal components as $K = K_s \otimes K_t$ with dimensions $L \times H \times 1$ and $1 \times 1 \times T$, respectively.
After each binary frame, the score is compared to a threshold $\eta$ and the pixel is disabled for the subsequent $\tau_H$ frames:
	$M(i,j,t') = 0$ for $\{t' | t+1 \leq t' \leq t+1+\tau_H\}$ if $S(i,j,t) > \eta$.
Observe in \cref{fig:snr_and_tuning}(b,c) that decreasing $\eta$ and increasing $\tau_H$ can be used to attain more aggressive inhibition with a larger fraction of photons being inhibited.
The binary rate of each pixel is estimated (using \cref{eq:num_det_inhibited}) as the ratio of detections to (active) measurements: $\widehat Y(i,j,t) = \sum\limits_{t'} F(i,j,t')/\sum\limits_{t'} M(i,j,t')$.
This calculation requires a record of the inhibition history which could be accumulated by an in-pixel counter or recreated in a downstream processor if all binary frames are read out.
\smallskip

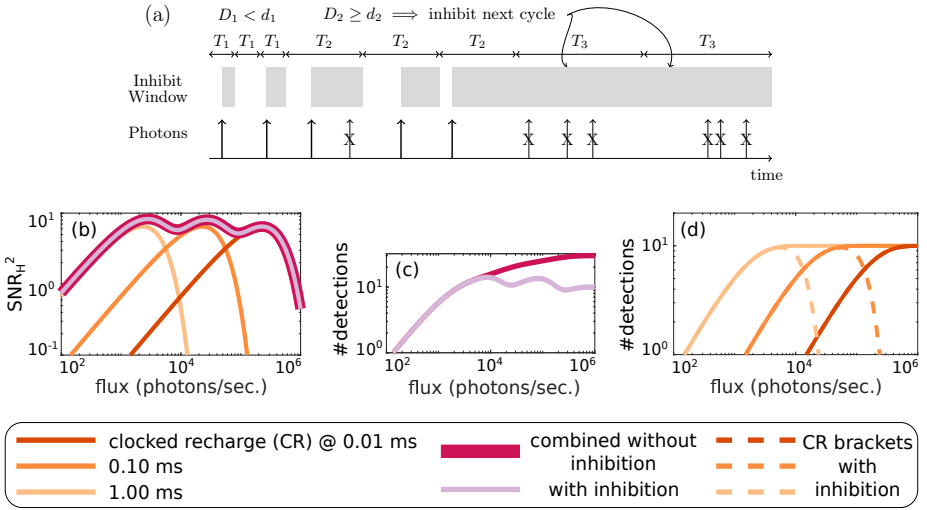
\begin{figure}[!bht]
\centering
	\begin{subfigure}{\linewidth}
	 \centering
	\begin{tabularx}{0.75\columnwidth}{ X }
		\small
		\adjustbox{max width=1\linewidth}{
	\begin{tikzpicture}[baseline=(current bounding box.center)]
		\newcommand\phht{0.75}
		\newcommand{\upph}[2]{\draw[->, thick] (#1, 0) -- (#1, #2)}
		\newcommand{\upphin}[2]{\draw[->, thin] (#1, 0) -- (#1, #2) node[midway]{X}}
		\newcommand{\expline}[3]{\draw[-, thin, color=white] (#1, 1.75) -- (#1, 2.25); \draw[<->] (#1, 2) -- (#1+#2, 2) node[midway,above] {#3}}

        \node (L) at (-1, 2.75) {\large (a)};

		\node (A) at (-1, 1.5) {Inhibit};
        \node (Z) at (-1, 1.2) {Window};

		\node (B) at (-1, 0.5) {Photons};
		\node (O) at (0,0) {};
		\node (D) at (5, 0) {};
		\node (D2) at (10.9, -0.3) {time};
		
		\draw[->] (0,0) to (11,0);
		\expline{0}{0.5}{$T_1$};
		\expline{0.5}{0.5}{$T_1$};
		\expline{1}{0.5}{$T_1$};
		\expline{1.5}{1.5}{$T_2$};
		\expline{3}{1.5}{$T_2$};
		\expline{4.5}{1.5}{$T_2$};
		\expline{6}{2.5}{$T_3$};
		\expline{8.5}{2.5}{$T_3$};
		
		\upph{0.25}{\phht};
		\draw[draw=none, fill=black!15] (0.25,1) rectangle (0.5,1.75);
		
		\upph{1.125}{\phht};
		\draw[draw=none, fill=black!15] (1.125,1) rectangle (1.5,1.75);
		
		\node (L1) at (0.75, 2.75) {$D_1<d_1$};	
		
		\upph{2}{\phht};
		\draw[draw=none, fill=black!15] (2,1) rectangle (3,1.75);
		\upphin{2.75}{\phht};

		\upph{3.75}{\phht};
		\draw[draw=none, fill=black!15] (3.75,1) rectangle (4.5,1.75);
		
		\upph{4.75}{\phht};
		\draw[draw=none, fill=black!15] (4.75,1) rectangle (6,1.75);
		
		\node (L2) at (4.5, 2.75) {$D_2 \ge d_2 \implies$ inhibit next cycle};	
		\draw [->] (7,2.75) to [out=30,in=150] (7,1.75);
		\draw [->] (7,2.75) to [out=30,in=70] (9,1.75);

		\upphin{6.25}{\phht};
		\draw[draw=none, fill=black!15] (6,1) rectangle (8.5,1.75);
		\upphin{7}{\phht};
		\upphin{7.5}{\phht};
		
		\upphin{9.75}{\phht};
		\draw[draw=none, fill=black!15] (8.5,1) rectangle (11,1.75);
		\upphin{10}{\phht};
		\upphin{10.5}{\phht};
	\end{tikzpicture}
 }   \\
	\end{tabularx}
	\end{subfigure}
	\begin{subfigure}{\linewidth}
		\centering
		\vspace{1em}
		\includegraphics{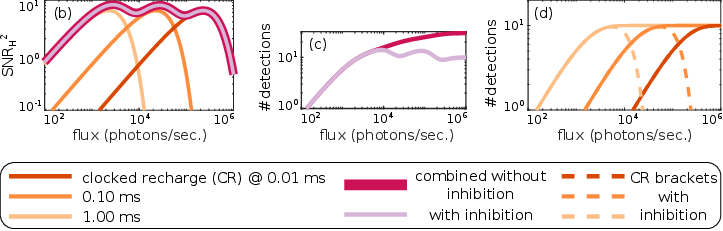}
	\end{subfigure}
	\caption{\textbf{Saturation look-ahead inhibition.}
		(a) When detections $D_i$ of cycle $i$ exceed a threshold, $d_i$, the photons arriving during the next recharge cycle, $i+1$, are inhibited.
		For this drawing $[d_1, d_2] = [3, 3]$.
		(b, c, d) Example with three recharge period settings and brackets of $W_i = 10$ measurements each.
		The saturation thresholds are set as $[d_1, d_2] = [7, 7]$.
		The brackets are combined linearly, weighted by $\snr_H^2$ \cite{gnanasambandamHDRImagingQuanta2020}.
		(b, c) With negligible effect on reconstructed SNR relative to conventional clocked-recharge, significantly fewer photon detections occur under the inhibition policy in high-flux settings.
		(d) The individual brackets are effectively disabled in flux regimes with low detection efficiency, and the detection rate closely tracks the original $\snr_H^2$ curve.
		Simulation details are in the supplement.}
	\label{table:sat_lookahead_policy}
\end{figure}

\noindent {\bf Example policies:}
\textit{1) Single-pixel dead time}: A spatial kernel $K_s$ of dimensions $1\times1$ may mimic the dead time of a passive SPAD without influence from neighboring pixels.
Unlike SPAD recharge generated dead time, the temporal kernel may be extended (e.g., $N$ frames long) to establish a rate threshold for inhibition with reduced quantization noise.
\textit{2) Local spatio-temporal averaging}: Single binary frames are inherently noisy.
An inhibition policy that calculates spatio-temporal averages to estimate the local photon rate may reduce the impact of noise on the inhibition pattern.
Another benefit, as shown in \cref{fig:snr_and_tuning}(b,c), is that the stochastic nature of binary frames smooths the distribution of measurements versus pixel flux.
Discontinuities (``dips'') in SNR versus photon flux are undesirable due to the potential for artifacts.
\textit{3) Edge enhancement:} Pixels may be inhibited if a local neighborhood has little spatial variation in photon rate.
This can be achieved, for instance, through a spatial filter $K_s$ in \cref{eq:inhibition_score} which acts like a Laplacian filter.
Such a strategy may enhance the fidelity of edges in the image while focusing fewer resources on regions with constant illumination.

\label{sec:lookahead}
\cref{table:sat_lookahead_policy} presents a second proposed inhibition policy called \textit{saturation look-ahead}.
This policy combines exposure brackets and calculation-based inhibition for a light-weight single-pixel inhibition policy.
This policy proceeds as a sequence of cycles (indexed by $i$) of binary frames where each binary frame within each cycle uses the same exposure time.
Cycle exposure times $T_i$ progressively increase ($T_1<T_2<T_3 ...$) so that measurements taken in an earlier cycle may predict low detection efficiency (near saturation) at longer exposure times and disable the pixel in these subsequent cycles.
The exposure level thresholds for inhibition would typically be set such that the number of photons detected at a given flux level tracks the $\snr_H^2$ (see Fig. \ref{table:sat_lookahead_policy}(b)), but may be adjusted further based on the relative importance of power consumption, sensing latency, and SNR in an actual application setting.

\section{Simulation-based Evaluation \label{sec:simulation_results}}
We use Monte Carlo simulations to generate sequences of grayscale binary frames from a dataset of RGB images \cite{arbelaezContourDetectionHierarchical2011}.
The inhibition policies evaluated extend the baseline inhibition generated by clocked recharge.
The inhibition score and patterns for various policies and tuning parameters are calculated from these binary frame sequences.
Once inhibition patterns are determined, performance is evaluated by tabulating detections, measurements, and image quality or vision task performance for each step in the sequence (see the supplement for details).

\begin{figure}[t!]
	\includegraphics[width=\linewidth]{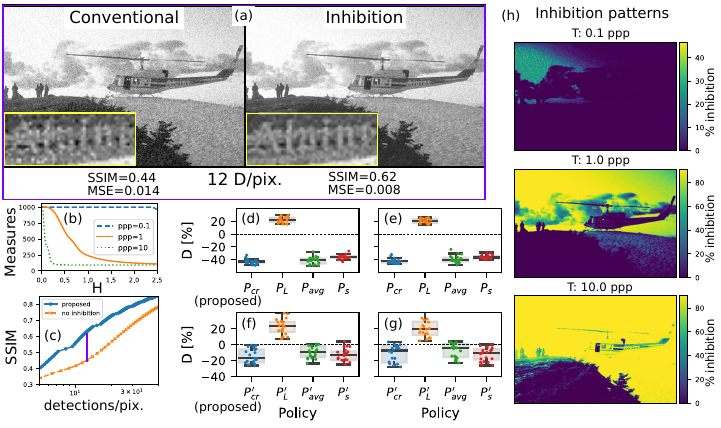}
	\caption{\textbf{Power-efficient static imaging via inhibition.}
		(a) Images from an exposure bracketing sequence (average exposure of $0.1,\, 1.0,\, 10.0$ photons per pixel (ppp)) using clocked recharge without (conventional) and with inhibition.
		(b) the distribution of measurements based on pixel intensity with $H$ at \SI{1.0}{ppp}.
		(c) image quality (SSIM) at equal detections/pixel.
		(d-g) Handcrafted policies are tested over 20 images and assessed by reductions in detections ($D$) at equal SSIM.
		The top row (d),(e) use exposure bracketing;
		(f),(g) use a single exposure of \SI{1.0}{ppp}.
		(d),(f) at SSIM=0.7 and
		(e),(g) at SSIM=0.8.
		The box shows quartiles with the center line at the median.
		The proposed policy, $P_{cr}$, is a 3\texttimes 3 spatial kernel that emphasizes the center pixel (\texttimes8) and includes the 8 neighbors (\texttimes1) (see the supplement for policy details).
		(h) average inhibition patterns for each exposure time.
		The top most pattern inhibits the brightest pixels only (maximum of $\sim$60\% inhibition, primarily in the sky).
		For the longest exposure time the inhibition pattern allocates measurements primarily to the darkest pixels.}
	\label{fig:box_whisker}
\end{figure}

\smallskip
\noindent \textbf{Spatio-Temporal Policies for Imaging:}
Handcrafted spatial kernels (3x3) were combined with an averaging temporal kernel of length 4 to form lightweight inhibition policies that allocate pixel measurements as described in earlier sections for improved image reconstruction.
\cref{fig:box_whisker} summarizes the simulation results.
\cref{fig:box_whisker}(d,e) display reduction in photon detections at equal structural similarity index measure (SSIM) \cite{wang2004image}, enabled by disabling bright pixels, for an exposure bracketing sequence.
Intensity estimates from each bracket are combined using $\snr^2$ weighting \cite{gnanasambandamHDRImagingQuanta2020} and then converted to a binary rate estimate at the center exposure level of \SI{1}{ppp}.
The proposed policy demonstrates an average reduction in detections of 42\% as compared to no inhibition.
\cref{fig:box_whisker}(f,g) evaluates a single exposure level (\SI{1.0}{ppp}) which is a more challenging scenario, yet the proposed policy still reduces detections by 14\% at SSIM=0.7.
See supplement S~3.4 for more examples and S~3.5 for simulations of high dynamic range images.

\begin{figure}[bt!]
	\centering
	\includegraphics[width=0.50\linewidth]{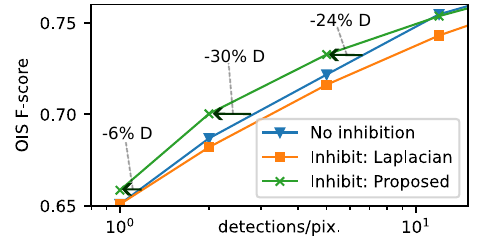}
	\caption{\textbf{Energy-efficient edge detection by inhibition}:
		Edge detection F-scores averaged over 19 images versus average detections/pixel.
		Horizontal arrows show reduction in detections enabled by inhibition at equal task performance.
		At and beyond 30 D/pix. no inhibition and the proposed policy are nearly equivalent and plateau at 200 D/pix.
		The edge detector returns $F=0.813$ on the original images of this set.}
	\label{fig:edge}
\end{figure}

\smallskip
\noindent \textbf{Edge Detection:}
The BSDS500 dataset with ground truth boundaries \cite{arbelaezContourDetectionHierarchical2011} was used to study energy-efficient edge detection via photon inhibition.
Binary rate images were processed by pre-trained holistically-nested edge detection (HED) \cite{xie15hed} with the resulting edge maps compared to ground truth by the structured edge detection toolbox \cite{dollarStructuredForestsFast2013}.
\cref{fig:edge} shows the optimal image scale (OIS) F-score versus the average detections per pixel.
Interpolated curves (not shown) allow for translating along horizontal lines of equal task performance to assess differences in avalanche energy.
At low photon counts the proposed edge-enhancing policy demonstrates a 30\% reduction in detections on single exposure time captures.
This policy extends the on-sensor calculation approach by calculating two scores, the conventional $3\times 3$ Laplacian \cite{wang2007laplacian} as $S_1$ and a $3\times 3$ averaging filter as $S_2$.
The final inhibition decision is the Boolean operation of these scores as $((\eta_1 {<} S_1 {<} \eta_2) \land (S_2 > \eta_3)) \lor (S_2 > \eta_4)$.
The Laplacian policy alone performs poorly as photons in dim regions with minimal spatial variation are inhibited -- disabling dim pixels is energy inefficient.
Complete policy descriptions are in the supplement.

\section{Experiments on Dynamic Scenes}
Many real-world scenes contain significant motion even with the high frame rate of a SPAD camera.
Burst reconstruction algorithms yield high-quality images from sequences of binary frames \cite{maQuantaBurstPhotography2020a,maBurstVisionUsing2023} -- we investigate their compatibility with the inhibited photon detection data.
In particular, we focus on the saturation look-ahead policy of Fig.~\ref{table:sat_lookahead_policy}(a), \emph{applied independently at every pixel}, and therefore an example of an adaptive single-pixel temporal policy.

To avoid losing salient information under motion, the inhibition policy must limit periods of extended dead time.
In the context of the saturation look-ahead policy this limits the total exposure length of the bracketing sequence, since we implicitly assume flux to be constant within each sequence.
In our experiments, we use a Fibonacci bracketing \cite{guptaFibonacci2013} sequence $T := \{1, 1, 2, 3, 5, 8, 13, 21\}$\footnote{Sequence $T' := \{1,1,1,3,3,3,8,8,25\}$ yielded similar results.
No extensive search over the policy space was performed.}, denoted in the units of a single minimal exposure time.
Every measured bracket is converted to a maximum-likelihood estimate of flux, which is then supplied to the quanta burst photography algorithm \cite{maQuantaBurstPhotography2020a}.
The flux estimator from bracketed measurements is described in detail in supplement Sec. S7.2.

\begin{figure}[t!]
	\centering
	\includegraphics[width=\textwidth]{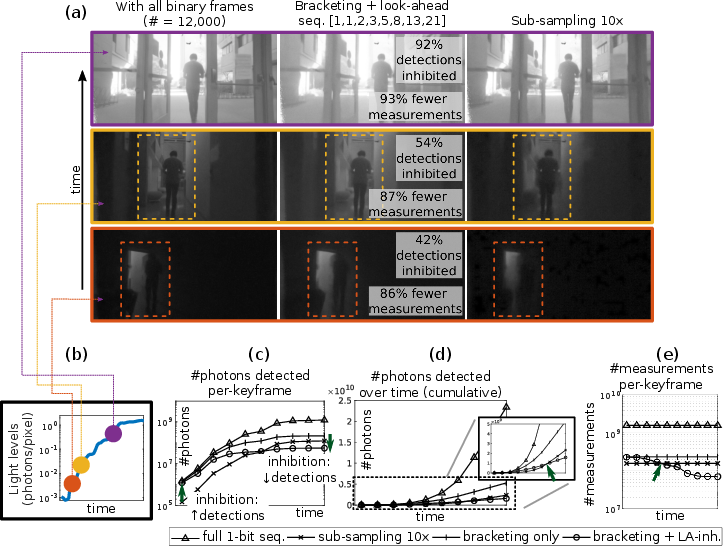}
	\caption{\textbf{Adaptive policies on video sequences enable stronger inhibition, preserve low-light details and, in bright-light, decouple flux and detection energy.}
		(a) Burst reconstructions \cite{maQuantaBurstPhotography2020a} for three keyframes with varying light levels: the top and bottom row differ by $\approx$ 7 stops.
		[Images show the detection rates $Y$ and are further gamma-compressed ($\gamma = 0.4$).
		Complete results included in the supplement.]
		Left column results are from the original binary frames without inhibition, and the right column after sub-sampling $10\times$ (a fixed $90\%$ inhibition). Middle column represents exposure bracketing combined with saturation look-ahead (see Fig.~\ref{table:sat_lookahead_policy}a for description).
		Under strong light (top row) the results are reasonable with both methods.
		However, plain sub-sampling loses details in lower light: notice the furniture and a person's outline in the middle \& bottom rows, respectively.
		Inhibition is instead adaptive to flux.
		(b) Average exposure level for each keyframe in the sequence.
		(c,d) Per-keyframe and cumulative detection counts -- inhibition ultimately results in fewer photons being detected over the whole sequence. (e) Number of measurements taken for each keyframe; reductions may be translated to energy savings during read-out. Plots in (c,d,e) are sub-sampled for clarity, and crossover points are marked by green arrows.}
	\label{fig:qbp_long_hdr_seq}
\end{figure}

\smallskip
\noindent \textbf{Results with the SwissSPAD2 sensor:}
The SwissSPAD2 sensor \cite{ulku512512SPAD2019} is a prototype SPAD pixel array that can produce binary frames at a rate up to $97{,}700$ FPS, with a resolution of $512{\times}256$.
In our experiments, we use binary frames captured directly (without inhibition) by the SPAD array as reference data and emulate on-sensor saturation look-ahead inhibition in software.
As a pre-processing step, measurements at hot pixels are replaced with their nearest neighbors.

Fig.~\ref{fig:qbp_long_hdr_seq} shows the results of burst reconstruction under three lighting conditions.
The raw data is a sequence of $>$580,000 binary frames with scene radiance increasing rapidly by orders of magnitude (Fig.~\ref{fig:qbp_long_hdr_seq}(b)), from $<$1 lux to $>$4{,}000 lux, measured separately with a light meter.
For each of $47$ equally-spaced keyframes, centered windows of $12{,}000$ binary frames are extracted and processed as described above.
Results for the full sequence can be found in the supplementary material.
A static inhibition policy of regular sub-sampling (dropping 9 out of every 10 frames) is also applied, which yields a fixed $90\%$ reduction in both measurements and photon detections under all lighting conditions.
Other sub-sampling factors are discussed in the supplement.

The top row in Fig.~\ref{fig:qbp_long_hdr_seq}(a) shows that under strong light a large fraction of photons ($> $90\%) can be inhibited through the saturation look-ahead policy and still result in good image quality after burst reconstruction, thus spreading photon detections over a longer period of time to reduce avalanche power.
Even simple sub-sampling yields good results in bright light, and may work well under controlled illumination conditions.
However, the images in the middle and bottom rows illustrate that this static inhibition policy results in excessive signal loss in lower light leading to a loss of details.
Sub-sampling may be enhanced by adjusting exposure time and/or the sub-sampling factor in response to global flux but cannot simultaneously optimize for different light levels within a single frame: over-exposed regions may have clipping artifacts and under-exposed regions are prone to motion blur from incorrect burst reconstruction.
The saturation look-ahead policy does adapt to local flux, and allocates relatively more detections to dim regions (see \cref{fig:overview}(d)).
It thus decouples detection energy from flux (Fig.~\ref{fig:qbp_long_hdr_seq}(c); cf. Fig. \ref{fig:overview}(a)), and results in cumulatively fewer detections than sub-sampling (Fig.~\ref{fig:qbp_long_hdr_seq}(d)) due to it being more aggressive in strong light.

Apart from photography, the burst-reconstructed images can be used in computer vision applications, where even stronger inhibition is possible depending on the noise- or blur-robustness of the vision algorithm.
Fig. \ref{fig:overview}(d) shows successful object detection with the YOLOv8 algorithm \cite{Jocher_Ultralytics_YOLO_2023} applied to a burst-reconstructed image, with approximately 95\% photon detections inhibited.

\FloatBarrier

\section{Limitations and Future Outlook \label{sec:discussion}}
\noindent \textbf{Implementation costs.}
Our focus in this paper is on reducing energy consumption due to avalanches.
Although avalanches contribute significant energy as compared to on-sensor computations~\cite{takatsuka36UmpitchSPAD2023, ota37W143dBDynamicRange1Mpixel2022, ardeleanComputationalImagingSPAD}, an important next step is to design a holistic model that includes the energy consumption of computations and readout (leveraging Eq. \ref{eq:meas_eff}).
Our in-pixel computations --- power-of-two multiplications which simplify to bit shifts, small spatio-temporal kernels no larger than 3\texttimes3\texttimes4 --- are designed to be lightweight.
Fortunately, computational SPAD imagers \cite{ardeleanComputationalImagingSPAD} with in-pixel memory and compute have recently been proposed, with 4\texttimes4 block of pixels having a 32-bit CPU and over 10~kbits of memory.
See the supplement for an estimate of the required circuitry. \smallskip

\noindent \textbf{More complete models.}
The noise model used for efficiency metrics only accounts for photon Poisson noise and the quantization noise of Bernoulli samples.
Expanding the noise model to include uncorrected pixel sensitivity variations, crosstalk, and afterpulsing may improve performance \cite{bianHighresolutionSinglephotonImaging2023} by allocating measurements with an awareness of practical sensor limitations.
Inhibition modulates the number of Bernoulli trials.
An unbiased estimator for data-dependent stopping of Bernoulli trials is known \cite{haldaneMETHODESTIMATINGFREQUENCIES1945}, yet is not applied in this work since it does not precisely match our situation.
Our simulations (SSIM) suggest that bias is less significant than image noise, yet future analytical studies are needed over a range of conditions. \smallskip

\noindent \textbf{Generalization to other tasks.}
To generalize the proposed approaches to a variety of vision tasks, task-specific quality metrics must be defined to pose optimization problems for each task.
As an example, for the image reconstruction task, Suppl. Sec. 3.1 shows an analytical optimization using the MSE metric when constrained by photon detections.
Other tasks could be approached similarly, yet may need to be optimized empirically.
\smallskip

\section*{Acknowledgments}
The authors acknowledge the Minnesota Supercomputing Institute (MSI) at the University of Minnesota for providing resources. A.I. was supported in part by NSF ECCS-2138471. S.G and M.G. were supported in part by NSF CAREER award 1943149, NSF award CNS-2107060, and Wisconsin Alumni Research Foundation (WARF).

\bibliographystyle{splncs04}
\bibliography{reflist}

\clearpage
\appendix

\section*{Supplementary Information}

\renewcommand{\figurename}{Suppl. Fig.}
\renewcommand{\tablename}{Suppl. Table}
\renewcommand{\thesection}{S \arabic{section}}
\renewcommand\theequation{S\arabic{equation}}
\setcounter{figure}{0}
\setcounter{table}{0}

\section{Photon Flux Values in Photometric Units}
Fig. 1(a) in the main text shows the rate of growth in camera power consumption as a function of photon flux in photons per second.
Unfortunately, due to the complications involved in direct conversion of radiometric photon flux into photometric quantities, it is not possible to provide exact numbers (in lux) for the photon flux.
For some intuition on the real-world lighting conditions that these photon flux values correspond to, we provide ``back of the envelope'' estimates in terms of lux levels using the following relationship:
$$
\text{illuminance (lux)} \approx \frac{hc}{\lambda}\frac{K\phi}{A}
$$
where $h=6.626\times 10^{-34}$ is the Planck's constant, $c=3\times 10^8$ is the speed of light in vacuum, $\lambda$ is the nominal visible wavelength of light which we assume is $\SI{555}{\nano\meter}$ at which the luminous efficacy of an ideal monochromatic light source is $K=\SI{683}{\lumen/\watt},$ $\phi$ is the incident photon flux (adjusted for the SPAD pixel's non-ideal quantum efficiency of 20\%), and $A$ is the effective pixel area (assuming a pixel pitch $\sim \SI{4}{\micro\meter}$ and fill factor of 10\%).
Plugging in the range of photon fluxes $10^3-10^6$ gives a range of lux levels from $<\SI{1}{\lux}$ to over $\SI{1000}{\lux}$.
These lux levels are approximately denoted by icons along the x-axis in Fig.~1(a) as $<\SI{1}{\lux}$ for a moonless night, $\sim1-\SI{10}{\lux}$ for twilight, $\sim\SI{100}{\lux}$ at sunrise or sunset, and $>\sim\SI{1000}{\lux}$ on a clear sunny day outdoors.

\section{Power consumption estimates}
Since we do not have a hardware prototype of our own, we make a very rough estimate here based on previously published works.
From Fig. 4.15 on pg. 97 of Andrei Ardelean's thesis \cite{ardeleanComputationalImagingSPAD}, computations in the UltraPhase imager effectively use $\sim$ 0.21 mW to constantly perform "MAC operations with data in registers" in a tight loop (= 1.19 mW total $-$ 0.98 mW "standby" power use).
Since UltraPhase has 12x24 = 288 pixels, this comes out to about 729 nW per pixel.
Energy consumption by avalanches is estimated in \cite{severini2021spad} as 11.6 pJ/avalanche, for a different SPAD sensor (not the SwissSPAD2).
This suggests that we should come out ahead if we can inhibit at least $\frac{729 nW}{11.6 pJ/detection} = 62{,}845$ detections/sec.
Under the parameters of Sec. S1 above, an example of this would be going from $90{,}000$ det./sec. (${\sim}230$ lux, mapping to daylight or office light) to $25{,}000$ det./sec. (effectively ${\sim}30$ lux, around dusk) --- still enough for a reasonable image with a SPAD sensor.
Therefore there exists a very plausible application setting where inhibition can make an impact.

The numbers above are clearly not specific to our sensor and computations.
The UltraPhase processing is reconfigurable and has a 32-bit wide arithmetic logic unit; whereas, inhibition processing would use fixed logic with smaller bit widths. As such we expect the above to be an over-estimate of the computation power.
We would also need to measure the avalanche energy expenditure of the SwissSPAD2 sensor instead of re-using the estimates from \cite{severini2021spad}.

\section{Spatio-temporal Policies for Static Imaging}
\subsection{Simulation Implementation Details}
In this section, the methods for the simulations of Section 6 of the main manuscript are described. Images from the BSDS500 dataset \cite{arbelaezContourDetectionHierarchical2011} were used to simulate binary-frames (specifically 20 images were randomly selected from the official test set).
This dataset was chosen due to the availability of ground truth edge maps. Images were gamma-decompressed using the sRGB to CIE XYZ transformation ($\gamma\approx2.2$) and converted to grayscale using the OpenCV color space conversion function (\textit{cvtColor} with COLOR\_BGR2YUV) to create a reference image.
For each reference image, 1,000 binary frames were simulated using Monte Carlo methods for each exposure time of interest and saved to disk. Exposure times are reported in units of the average number of photon arrivals per pixel (ppp), since absolute radiometric quantities are not available.

For static imaging, inhibition policy simulations were run for each exposure time separately.
Once inhibition patterns are found for each frame index $t$, the cumulative detections and measurements are calculated for each frame index. This approach allows for extraction of performance metrics and images at a continuous range of average detections per pixel by selecting the number of accumulated binary frames.
For exposure bracketing simulations an HDR reconstruction was generated at each frame index using $\snr^2$ weighting \cite{gnanasambandamHDRImagingQuanta2020}.
Metrics of SSIM \cite{wang2004image} and mean squared error (MSE) were calculated on binary rate images for the accumulated binary frames with and without inhibition at each frame index using the original image as the reference.

\subsection{Assessing Inhibition}
Pixels that are inhibited are known at the beginning of a frame.
An inhibited pixel is insensitive to photon arrivals and, as such, does not consume (avalanche) power when a photon converts.
Inhibition is expected to be implemented by lowering or keeping the pixel SPAD bias below the threshold voltage for an avalanche.
A pixel that is not inhibited measures either a '0' (if no photons arrived) or a '1'.
The avalanche energy is assumed the same for one and more than one photon in a single frame, which has been demonstrated in hardware \cite{morimotoMegapixelTimegatedSPAD2020}.
To assess the energy efficiency, measurement efficiency, and energy reduction enabled by inhibition we track the number of measurements ($W$) at each pixel and total measurements ($W_T$), and similarly the number of detections ($D$) for each pixel and total detections ($D_T$).
These quantities are defined as follows:
\begin{align}
	W(i,j) &= \sum_{t=0}^{t=N-1} (1 - M(i,j,t)) \\
	W_T &= \sum_{i,j} W(i,j) \\
	D(i,j) &= \sum_{t=0}^{t=N-1} F(i,j,t) = \sum_{t=0}^{t=N-1}  (1 - M(i,j,t)) Y(i,j,t) \\
	D_T &= \sum_{i,j} D(i,j),
\end{align}
where $N$ denotes the number of binary frames.
Measurements $W(i,j)$ are the total number of frames during which pixel $(i,j)$ was not inhibited (i.e., the inhibition pattern $M(i,j)=0$).
The number of measurements may correlate with the readout energy if an unconventional readout architecture, such as token passing \cite{gabrielliFastReadoutPixel2008} or asynchronous event readout \cite{berkovichScalable20202015}, is combined with the inhibition pattern.
Detections $D(i,j)$ are the total number of frames during which a photon was detected by pixel $(i,j)$ when enabled.
The number of detections tracks the total avalanche energy consumed by that pixel.
Suppl. Table~\ref{table:assess} summarizes relevant parameters used for evaluating different inhibition policies.

\begin{table*}[!htbp]
	\centering
	\renewcommand{\arraystretch}{1.1}
	\begin{tabular}{ |l|c|c| }
		\hline
		Description & Variable & Values / Units \\
		\hline
		Photon flux & $\phi (i,j,t)$ & photons/s \\
		Binary frame exposure time & $T$ & s \\
		Exposure & $ H(i,j,t) = \phi(i,j,t)T$ &  photons \\
		Inhibition pattern (disabled = 1) & $M(i,j,t)$ & 0/1  \\
		Incident binary frame & $Y(i,j,t) \sim \mathrm{Bernoulli}(1 - e^{-H(i,j,t)})$ & 0/1\\
		Binary frame (after inhibition) & $F(i,j,t) = Y(i,j,t) \cdot (1-M(i,j,t)) $ & 0/1  \\
		Binary rate estimate & $\widehat{Y}(i,j) = \sum\limits_t \frac{F(i,j,t)}{M(i,j,t)} $  & [0,1] \\
		Explicitly inhibited photons & $I(i,j,t)= Y(i,j,t) \cdot M(i,j,t)$ & 0/1 \\
		Total photon detections & $D_T = \sum\limits_t D(i,j) = \sum\limits_t F(i,j,t)$ & photons \\
		Fraction of (possible) photons inhibited & $I_F = \frac{\sum I(i,j,t)}{\sum Y(i,j,t)}$ & [0,1]\\
		\hline
	\end{tabular}
	\renewcommand{\arraystretch}{1}
	\caption{Relevant quantities for assessing inhibition policies.
		Pixels are indexed by $i$ and $j$ while $t = 0,1,...,N-1$ is the discrete frame number.
		Explicitly inhibited photons are due to the inhibition pattern itself and not clocked recharge policy.}
	\label{table:assess}
\vspace{-1.5em}
\end{table*}

\subsection{Details of Imaging Policies} \label{sec:imaging_policies}
Sec.~5 and Figures~5,~6 in the main text describe static inhibition policies that use spatio-temporal information to compute inhibition patterns.
The aggressiveness of these policies is controlled through two parameters $\eta$ and $\tau_H$:
lower values of $\eta$ and higher values of $\tau_H$ can be used to reduce the number of measurements, and hence reduce the total avalanche energy consumption.
The policies shown below are the best performing combinations of $\eta$ and $\tau_H$ on exposure bracket captures shown in Fig.~7 (main text) for each of the four spatial policies presented.
Policies are designed so that multiplications can be implemented using bit shifts (powers of two) for ease of future in-pixel hardware implementation.
\vspace{-0.3em}
\begin{align*} 
	\mathbf{P_{cr}}: \;
	K_s &=
	\begin{bmatrix}
		1 & 1 & 1 \\
		1 & 8 & 1 \\
		1 & 1 & 1
	\end{bmatrix}, \; \; \; \;
	K_T =
	\begin{bmatrix}
		1&1&1&1 \\
	\end{bmatrix}, \;
	\eta = 12, \;
	\tau_H = 32; \;
   \textrm{``Center + ring''} \\
 \mathbf{P_{L}}:  \;
	K_s &=
	\begin{bmatrix}
		1 & 1 & 1 \\
		1 & -8 & 1 \\
		1 & 1 & 1
	\end{bmatrix}, \;
	K_T =
	\begin{bmatrix}
		1 & 1 & 1 & 1 \\
	\end{bmatrix}, \;
	\eta =  24, \;
	\tau_H = 4; \;
 \textrm{``Laplacian''} \\
 \mathbf{P_{avg}}:
	K_s &=
	\begin{bmatrix}
		1 & 1 & 1 \\
		1 & 1 & 1 \\
		1 & 1 & 1
	\end{bmatrix}, \; \; \; \;
	K_T =
	\begin{bmatrix}
		1 & 1 & 1 & 1 \\
	\end{bmatrix}, \;
	\eta = 6, \;
	\tau_H = 32, \;
  \textrm{``Average''} \\
	\mathbf{P_{s}}: \;
	K_s &=
	\begin{bmatrix}
		0 & 0 & 0 \\
		0 & 1 & 0 \\
		0 & 0 & 0
	\end{bmatrix}, \; \; \; \;
	K_T =
	\begin{bmatrix}
		1 & 1 & 1 & 1 \\
	\end{bmatrix}, \;
	\eta = 2, \;
	\tau_H = 32; \textrm{``Single pixel''}
\end{align*}
\noindent The policies described below (and annotated with a $\mathbf{'}$) are the best performing combinations of $\eta$ and $\tau_H$  on single-exposure time captures in Fig. 7 (sub-figures f and g) of the main manuscript for each of the four spatial policies presented.
\begin{align*} 
	\mathbf{P'_{cr}}: \;
	K_s &=
	\begin{bmatrix}
		1 & 1 & 1 \\
		1 & 8 & 1 \\
		1 & 1 & 1
	\end{bmatrix},  \; \; \; \;
	K_T =
	\begin{bmatrix}
		1 & 1 & 1 & 1 \\
	\end{bmatrix}, \;
	\eta = 12, \;
	\tau_H = 4; \;
 \textrm{``Center + ring''} \\
	\mathbf{P'_{L}}:
	K_s &=
	\begin{bmatrix}
		1 & 1 & 1 \\
		1 & -8 & 1 \\
		1 & 1 & 1
	\end{bmatrix}, \;
	K_T =
	\begin{bmatrix}
		1 & 1 & 1 & 1 \\
	\end{bmatrix}, \;
	\eta =  24, \;
	\tau_H = 4; \;
    \textrm{``Laplacian''} \\
	\mathbf{P'_{avg}}: \;
	K_s &=
	\begin{bmatrix}
		1 & 1 & 1 \\
		1 & 1 & 1 \\
		1 & 1 & 1
	\end{bmatrix},  \; \; \; \;
	K_T =
	\begin{bmatrix}
		1 & 1 & 1 & 1 \\
	\end{bmatrix}, \;
	\eta = 12, \;
	\tau_H = 4; \;
 \textrm{``Average''} \\
	\mathbf{P'_{s}}:, \;
	K_s &=
	\begin{bmatrix}
		0 & 0 & 0 \\
		0 & 1 & 0 \\
		0 & 0 & 0
	\end{bmatrix},  \; \; \; \;
	K_T =
	\begin{bmatrix}
		1 & 1 & 1 & 1 \\
	\end{bmatrix}, \;
	\eta = 2, \;
	\tau_H = 8; \;
 \textrm{``Single pixel''} \\
\end{align*}
\noindent Inhibition policies for static imaging were studied by simulations at parameters values of $\eta = [2,6,12,24]$ and $\tau_H = [4,8,16,32]$. A more extensive search was not attempted due to computation time and disk usage.

\newpage
\subsection{Additional Static Image Simulation Results}
\begin{figure}[hb!]
\centering
	\includegraphics[width=\linewidth]{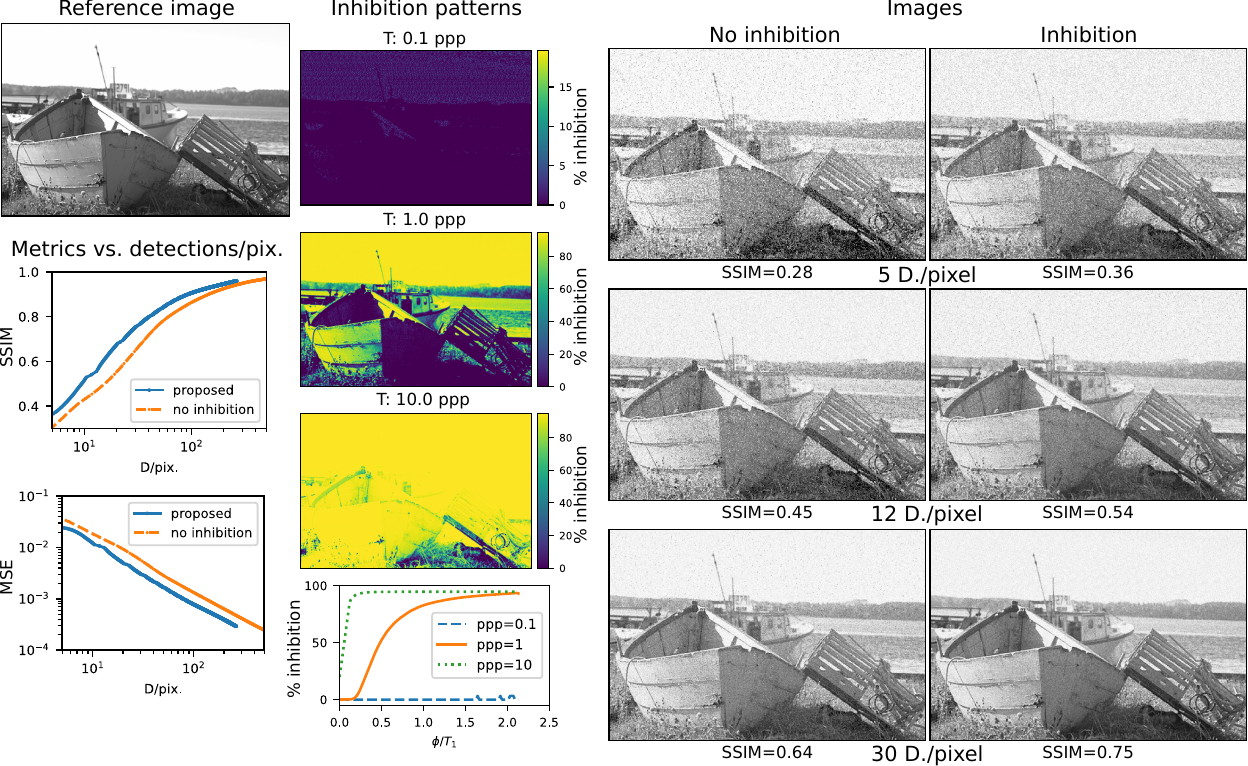} \vspace{-0.2in}
	\caption{\textbf{Power-efficient static single-photon imaging via inhibition.}
		A reference image (BSDS500: 393035) displayed in the top left is captured using a bracket of three exposure times with 1,000 binary frames for each exposure time.
		The second column displays the resulting average inhibition patterns for each exposure time.
		The top most pattern from the shortest exposure time modestly inhibits and does so at the brightest pixels only.
		The inhibition pattern of the longest exposure time allocates most measurements to the darkest areas of the scene (in the shadows to the right of boat in the the foreground).
		The bottom chart summarizes the inhibition patterns using smoothed curves of the inhibition percent versus the flux of each pixel for each of the three exposure times (Lowess filter with a fraction of 1/5).
		The right-most columns show binary rate images using gamma compression ($\gamma = 0.4$) without (left) and with (right) inhibition at equal average detections per pixel.
		Detections increase moving down with averages of 5, 12, and 30 detections per pixel shown.
		Image quality metrics versus detections per pixel are summarized in the center and bottom of the left most column.
	\vspace{-0.1in}}
	\label{fig:static_collage1}
\end{figure}

\begin{figure}[t!]
	\centering
	\includegraphics[width=\linewidth]{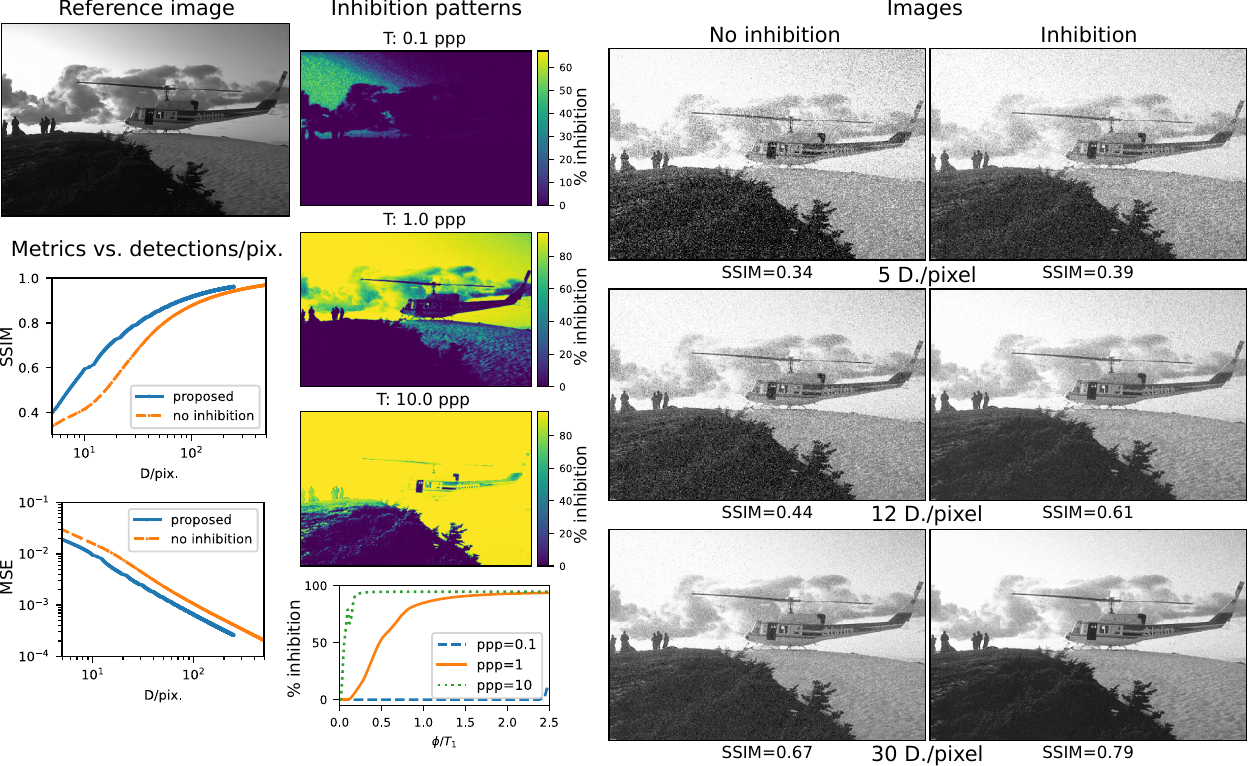}
	\caption{\textbf{Power-efficient static single-photon imaging via inhibition.}
		A reference image (BSDS500: 179084) displayed in the top left is captured using a bracket of three exposure times with 1,000 binary frames for each exposure time.
		The second column displays the resulting average inhibition patterns for each exposure time.
		The top most pattern from the shortest exposure time only modestly inhibits and does so at the brightest pixels only (maximum of $\sim$60\% inhibition, primarily in the sky).
		The inhibition pattern of the longest exposure time allocates most measurements to the darkest areas of the scene (the hilltop and the dark areas of the helicopter).
		The bottom chart summarizes the inhibition patterns using smoothed curves of the inhibition percent versus the flux of each pixel for each of the three exposure times (Lowess filter with a fraction of 1/5).
		The right-most columns show binary rate images using gamma compression ($\gamma = 0.4$) without (left) and with (right) inhibition at equal average detections per pixel.
		Detections increase moving down with averages of 5, 12, and 30 detections per pixel shown.
		Image quality metrics versus detections per pixel are summarized in the center and bottom of the left most column.}
	\label{fig:static_collage2}
\end{figure}

\begin{figure}[t!]
	\centering
	\includegraphics[width=\linewidth]{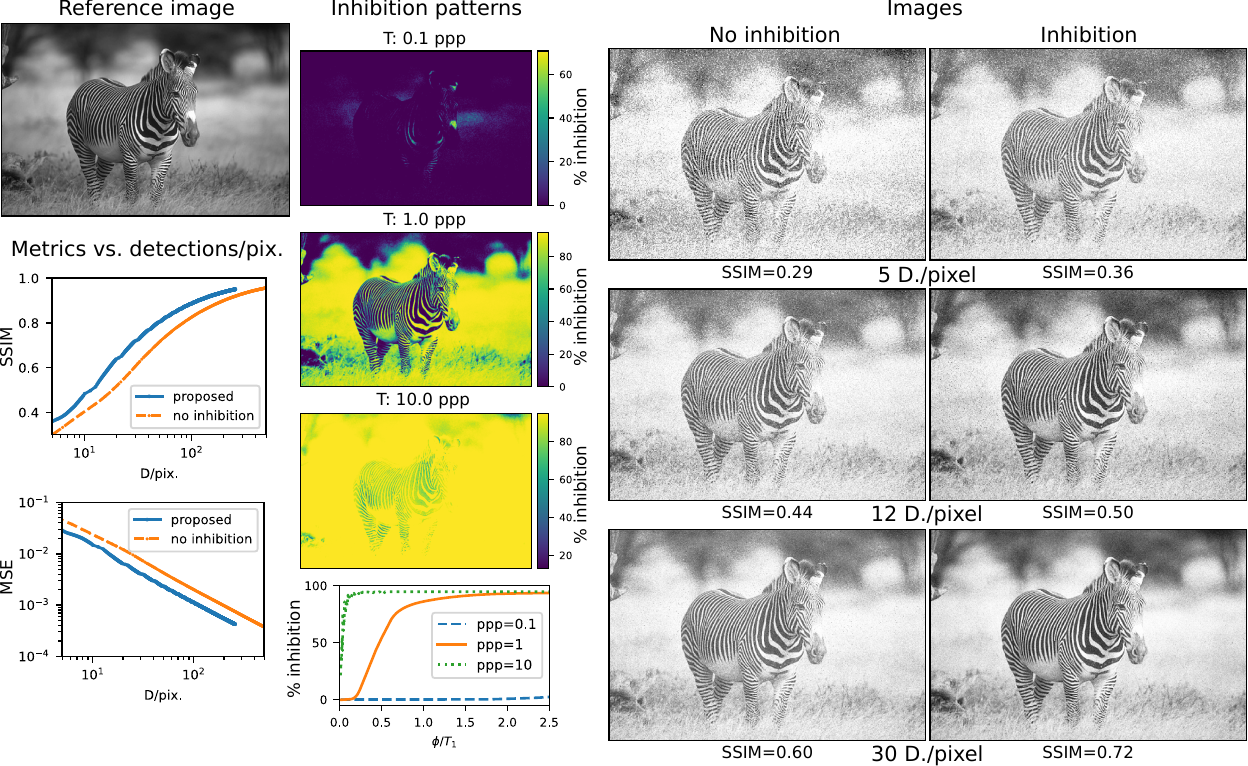}
	\caption{\textbf{Power-efficient static single-photon imaging via inhibition.}
		A reference image (BSDS500: 130066) displayed in the top left is captured using a bracket of three exposure times with 1,000 binary frames for each exposure time.
		The second column displays the resulting average inhibition patterns for each exposure time.
		The top most pattern from the shortest exposure time only modestly inhibits and does so at the brightest pixels only.
		The inhibition pattern of the longest exposure time allocates most measurements to the darkest areas of the scene (the dark stripes of the zebra).
		The bottom chart summarizes the inhibition patterns using smoothed curves of the inhibition percent versus the flux of each pixel for each of the three exposure times.
		(Lowess filter with a fraction of 1/5).
		The right-most columns show binary rate images using gamma compression ($\gamma = 0.4$) without (left) and with (right) inhibition at equal average detections per pixel.
		Detections increase moving down with averages of 5, 12, and 30 detections per pixel shown.
		Note the improved contrast of the image captured using an inhibition policy in the bottom row (30 D./pix).
		Image quality metrics versus detections per pixel are summarized in the center and bottom of the left most column.}
	\label{fig:static_collage3}
\end{figure}

Suppl. Figs.~\ref{fig:static_collage1}, \ref{fig:static_collage2}, and \ref{fig:static_collage3} expand upon the results of Fig. 5 of the main text to show inhibition patterns and resulting images at three levels of average detections per pixel using the $P_{cr}$ policy described above with $\eta=12$ and $\tau_H=32$.

\FloatBarrier

\begin{figure}[htb!]
\centering
	\includegraphics[width=\linewidth]{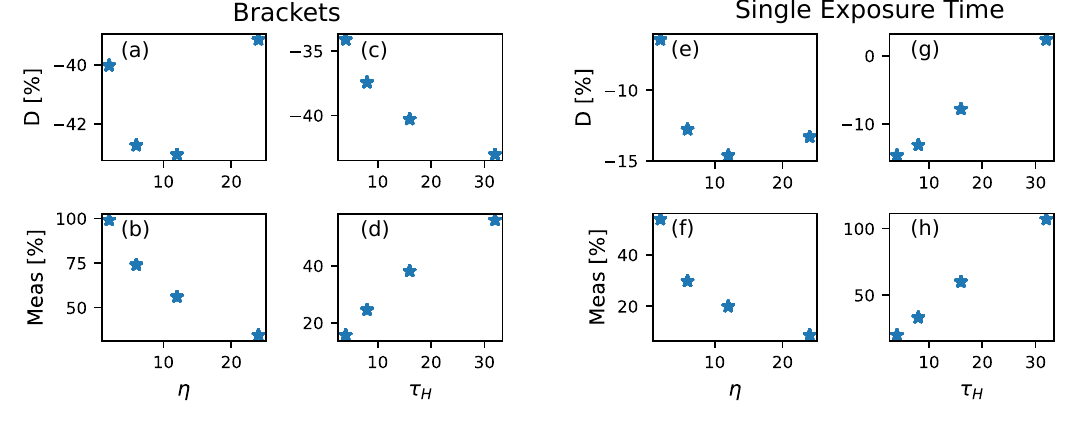}
	\caption{\textbf{Inhibition tuning parameters tradeoff detections and measurements.}
		For the $P_{cr}$ (exposure brackets, left) and $P'_{cr}$ (single exposure, right) policies percent change in detections (D) and measurements at SSIM=0.7 with one parameter varied.
		(a) D\% (as compared to without inhibition) versus the inhibition threshold $\eta$ at a constant holdoff time ($\tau_H=32$).
		A negative value, as in (a), indicates that the inhibition policy required fewer detections for equal SSIM.
		Notice in (b) how measurements (as \% of total possible) increase (measurement efficiency degrades) at more aggressive inhibition thresholds (smaller $\eta$).
		(c) Shows the impact of the holdoff time ($\tau_H$) at a constant threshold of $\eta=12$.
		(e, f, g, h) show the same for a single exposure time capture.
		For (e, f) $\tau_H = 4$ and for (g, h) $\eta=12$.}
	\label{fig:parameter_tuning}
\end{figure}

Suppl. Fig.~\ref{fig:parameter_tuning} shows the average percent change from no inhibition to inhibition in detections (top row) and measurements (bottom row) at equal image quality versus two parameters of the $P_{cr}$ proposed inhibition policy.
These charts demonstrate the balance of detection efficiency and measurement efficiency.
Efficiency improvements via inhibition correspond to negative values.
For the exposure bracket scenario, the improvement in detection efficiency by more aggressive inhibition ($\tau_H{}\uparrow$) shown in (c) increases the measurements, and hence degrades measurement efficiency (d).
As seen in (e, f, g, h) a single exposure time policy performs best with less aggressive inhibition ($\tau_H=4$) since frames with a shorter exposure time are not available to fill in missing information for the brightest and aggressively inhibited pixels.
Yet, single exposure time policies still reduce detections by nearly 15\%.

\newpage
\subsection{High Dynamic Range Simulation Results}
\begin{figure}[hb!]
	\centering
	\includegraphics[width=\linewidth]{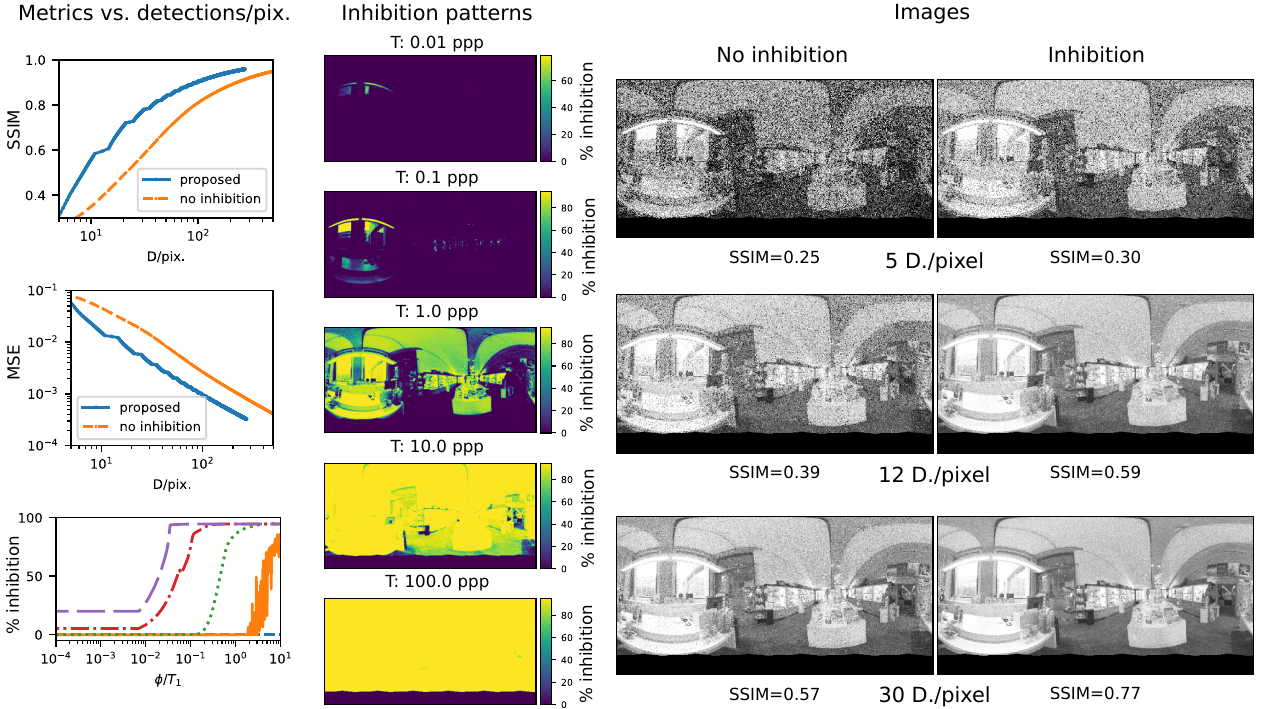}
	\caption{\textbf{Power-efficient static single-photon imaging via inhibition.}
		A reference image (9C4A0599; \SI{143.5}{dB} dynamic range) is captured using a bracket of five exposure times with 1,000 binary frames for each exposure time.
		The second column displays the resulting average inhibition patterns for each exposure time.
		The top most pattern from the shortest exposure time modestly inhibits and does so at the brightest pixels only.
		The inhibition pattern of the longest exposure time allocates most measurements to the darkest areas of the scene.
		The bottom chart in the leftmost column summarizes the allocations of measurements using smoothed curves of the inhibition percent versus the flux of each pixel for each of the three exposure times (Lowess filter with a fraction of 1/5).
		The right-most columns show binary rate images using gamma compression ($\gamma = 0.4$) without (left) and with (right) inhibition at equal average detections per pixel.
		Detections increase moving down with averages of 5, 12, and 30 detections per pixel shown.
		Image quality metrics versus detections per pixel are summarized in the leftmost column.
	\vspace{-0.1in}}
	\label{fig:hdr_static_collage1}
\end{figure}

\begin{figure}[hb!]
\centering
	\includegraphics[width=1\linewidth]{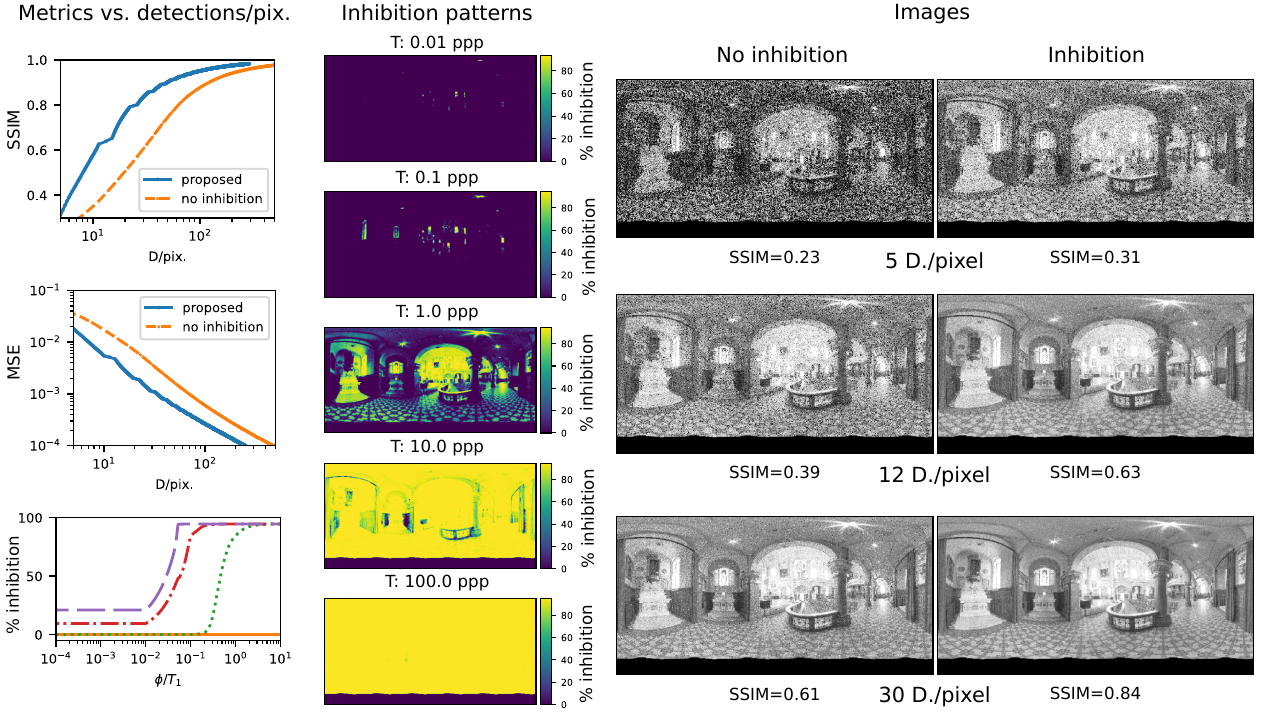}
	\caption{\textbf{Power-efficient static single-photon imaging via inhibition.}
		A reference image (AG8A7597; \SI{173.6}{dB} dynamic range) is captured using a bracket of five exposure times with 1,000 binary frames for each exposure time.
		The second column displays the resulting average inhibition patterns for each exposure time.
		The top most pattern from the shortest exposure time modestly inhibits and does so at the brightest pixels only.
		The inhibition pattern of the longest exposure time allocates most measurements to the darkest areas of the scene.
		The bottom chart summarizes the allocation of measurements using smoothed curves of the inhibition percent versus the flux of each pixel for each of the three exposure times (Lowess filter with a fraction of 1/5).
		The right-most columns show binary rate images using gamma compression ($\gamma = 0.4$) without (left) and with (right) inhibition at equal average detections per pixel.
		Detections increase moving down with averages of 5, 12, and 30 detections per pixel shown.
		Image quality metrics versus detections per pixel are summarized in the leftmost column.
	\vspace{-0.1in}}
	\label{fig:hdr_static_collage2}
\end{figure}

\textbf{Score-based inhibition}:
We also assessed the impact of scored-based inhibition to high dynamic range images from the Laval indoor HDR dataset \cite{gardnerLearningPredictIndoor2017a}.
Suppl. Figs.~\ref{fig:hdr_static_collage1} and \ref{fig:hdr_static_collage2} summarizes the results from the $P_{cr}$ policy described above with $\eta=12$ and $\tau_H=32$.
To accommodate the wide range of illumination in these images the simulations used five logarithmically spaced exposure times (in steps of \texttimes10).
Before simulation, the images were reduced in size by \texttimes4 along both dimensions using openCV resize with the default bilinear interpolation method to decrease the time required for simulation.
When capturing high dynamic range scenes, inhibition allows for a wide range of exposure times to efficiently measure bright and dim pixels with a reduced increase in avalanches at the brightest pixels.
For these experiments the exposure time sequence was not carefully explored.
Future work could optimize the sequence of exposure times in concert with the inhibition policy while using a more holistic energy cost model.

\FloatBarrier

\noindent \textbf{Saturation look-ahead inhibition}:
We estimated the benefits of saturation look-ahead inhibition analytically using the 10 images with the widest dynamic range in the Laval indoor HDR dataset (up to \SI{178}{dB} DR) \cite{gardnerLearningPredictIndoor2017a}.
The specific policy uses an exponential bracketing scheme with 5 exposure times scaled as $T_{n+1} = 5T_{n}$, and 10 measurements are taken with each one.
The thresholds to inhibit subsequent exposures under the saturation look-ahead policy were set as $D = \{6,6,6,6\}$, to ensure that on average, the flux range crossing them has a probability of photon detection of 0.99 or greater in the next (longer) exposure, representing (near) saturation.
That is to say, taking $T_1<T_2$, the detection rate $Y_1$ at the exposure time of $T_1$ that corresponds to a near-saturating detection rate of $Y_2$ at exposure time $T_2$ is computed as:
\begin{equation}
	Y_1 = 1 - e^{\frac{T_1}{T_2}\log(1-Y_2)} = 1 - e^{0.2\times\log(1-0.99)} \approxeq 0.60  = 6/10.
\end{equation}
The analytical results calculated using a subset of pixels from each of the images show that exposure brackets inhibit 90.6\% of detections as compared to a minimum exposure time of equal observation length.
The saturation look-ahead policy further reduces detections by an average of -38.4\% as compared to bracketing alone.
In total, averaged over the 10 images, look-ahead inhibition with exposure bracketing inhibits 94.0\% of the detections.
The images studied were 9C4A6135, AG8A3343, AG8A2979, AG8A5920, AG8A7597, AG8A6813, 9C4A3821, 9C4A3335, 9C4A1696, and 9C4A0599.

\FloatBarrier
\subsection{Details of Edge Detection Policies}
A high performing edge detection policy is presented in Section 6.2 and Fig. 6 of the main manuscript.
This policy calculates a score from the 3\texttimes3 Laplacian ($S_1$) and a 3\texttimes3 average filter ($S_2$).
The final inhibition decision is the Boolean operation of these scores as
	$((\eta_1 {<} S_1 {<} \eta_2) \land (S_2 > \eta_3)) \lor (S_2 > \eta_4)$.
The thresholds for the Laplacian score $S_1$ are $\eta_1 = -12$, $\eta_2 = 12$ which detects regions of minimal spatial variation.
To inhibit based on minimal spatial contrast, we additionally require that the average score exceeds a modest threshold, $\eta_3 = 4$, so that dim neighborhoods are not inhibited.
Finally, independent of the Laplacian calculation, the pixel is inhibited if the average score is excessive, $\eta_4=16$.
For this policy, $\tau_H=16$.
As in the policies for static imaging, the temporal kernel was
	$K_T =	\begin{bmatrix}
				1 & 1 & 1 & 1 \\
			\end{bmatrix}$.

\newpage
\section{Photon Efficiency and Metrics}
\subsection{Oracle Measurement Allocation \label{sec:oracle_meas_alloc_Y}}

This supplemental section motivates the assertion that an optimal measurement allocation that minimizes image MSE allocates measurements in proportion to $\sqrt{1-Y}$.
Photon inhibition enables an unequal distribution of measurements among the pixels of the sensor.
To guide the design of inhibition policies presented in the main manuscript we considered two questions.
How should a fixed number of detections be allocated among pixels of a single-photon sensor to optimize the image mean squared error (MSE)?
By what amount can image metrics be improved by measurement allocation policies when total detections are constrained?

The variance of the binary rate estimate of pixel $i$ with binary rate $Y_i$ and allocated $W_i$ measurements is
\begin{equation}
	\sigma_{Y_i}^2 = \frac{1}{W_i} Y_i(1-Y_i) \label{eqn:variance}.
\end{equation}
The MSE is the sum of the variances over all pixels $P$.
A minimum of the MSE is found by differentiating with respect to $W_i$ and constraining the total detections to $D_T = \sum\limits_i^P Y_i W_i$.
The optimal measurements allocated to pixel $i$ for this minimum MSE is then
\begin{align}
	W_{i}^{*} &= \frac{D_T}{\sqrt{Y_i}} \left( \frac{\sqrt{Y_i(1-Y_i)}}{\sum\limits_{j=1}^{j=P} Y_j\sqrt{1-Y_j}} \right).
	\label{eqn:optimal-MSE}  
\end{align}
\Cref{eqn:optimal-MSE} shows that when constrained by total detections, total image MSE is minimized when measurements are allocated in proportion to $\sqrt{1-Y}$.
This analytical approach requires perfect knowledge of the photon flux at each pixel, a non-causal ``oracle'', and is thus an upper bound on the improvements enabled by photon inhibition.
Similar optimal allocation policies may be developed for a scenario with a constraint on measurements rather than detections.
These allocations were derived separately and then later very similar allocations were found in the work of Medin et al. which derives optimal stopping rules for active imaging systems designed to estimate Bernoulli parameters \cite{medinBinomialNegativeBinomial2019}.

To validate \cref{eqn:optimal-MSE} we simulated the same 20 BSDS500 images as used in the main manuscript at a single effective exposure time of an average of 1 photon per pixel.
To prevent unbounded allocations, pixels were forced to a minimum rate of $Y=0.01$ and a maximum rate of $Y=0.99$.
The variance of each pixel was calculated as \cref{eqn:variance} and the image MSE was evaluated as the average of the variances of all pixels.
Two measurement allocations were evaluated.
The first approach allocated measurements equally between all pixels (``Uniform''); the second approach allocated measurements as defined in \cref{eqn:optimal-MSE} (``MSE optimal'').
With the image variance calculated from the ground truth binary rate and the measurements allocated to each pixel, a noise corrupted image was generated by adding noise from Gaussian distributed samples at each pixel to the ground truth image.
The noise image was then clipped to between [0,1].
This noise image was used to assess image quality metrics, such as SSIM, by comparison to the reference image.
Supp. Table~\ref{table:oracle} demonstrates these simulated results using the oracle ``MSE optimal'' measurement allocations.
The simulations suggest that the inhibition policy results presented in the main manuscript ($\sim$15\% detection reduction) approach the limits established by an oracle allocation for these specific images.
Potential sources of discrepancy include that an analytical allocation was only available for image MSE and because the oracle simulations added Gaussian distributed noise.

\begin{landscape}
\begin{table}[h!]
\centering
\begin{tabular}{c| l|| l|l || l|l}
\textbf{Avg. D/pix.} & \textbf{Allocation method} & \textbf{MSE$\downarrow$}     & \textbf{SSIM$\uparrow$}  & \textbf{equal MSE: D \% }& \textbf{equal SSIM: D \%} \\
\hline
\hline
\multirow{2}{*}{5}                    & Uniform           & 5.23$\times 10^{-3}$ $\pm$ 2.89$\times 10^{-3}$ & 0.696  $\pm$ 0.117 &                 &                  \\
                     & MSE optimal       & 4.47$\times 10^{-3}$ $\pm$ 2.64$\times 10^{-3}$ & 0.724 $\pm$ 0.116 & -10.0 $\pm$ 7.8          & -13.7 $\pm$ 6.87           \\
\hline
\multirow{2}{*}{12}                   & Uniform           & 2.18$\times 10^{-3}$ $\pm$ 1.21$\times 10^{-3}$ & 0.815  $\pm$ 0.084 &                 &                  \\
                     & MSE optimal       & 1.86$\times 10^{-3}$ $\pm$ 1.10$\times 10^{-3}$ & 0.837  $\pm$ 0.081 & -10.0 $\pm$ 7.8         & -13.4 $\pm$ 7.09           \\
\hline
\multirow{2}{*}{30}                 & Uniform           & 8.71$\times 10^{-4}$ $\pm$ 4.82$\times 10^{-4}$ & 0.905  $\pm$ 0.048 &                 &                  \\
                     & MSE optimal       & 7.45$\times 10^{-4}$ $\pm$ 4.40$\times 10^{-4}$ & 0.918  $\pm$  0.046 & -10.0 $\pm$ 7.8          & -12.8  $\pm$ 7.44          \\
\hline
\multirow{2}{*}{100}                  & Uniform           & 2.61$\times 10^{-4}$ $\pm$ 1.32$\times 10^{-4}$ & 0.967  $\pm$ 0.018 &                 &                  \\
                     & MSE optimal       & 2.23$\times 10^{-4}$ $\pm$ 1.45$\times 10^{-4}$ & 0.971  $\pm$ 0.017 & -10.0 $\pm$ 7.8          & -12.2 $\pm$ 7.43
\end{tabular}
	\caption{Average results from simulations of 20 images using an oracle allocation policy that is optimized for MSE as compared to a uniform distribution of pixel measurements.
		In alignment with the main manuscript, the last two columns show the percent change in detections for equal MSE and SSIM enabled by the oracle allocation policy as compared to a uniform allocation of measurements.
		$\pm$ indicates the standard deviation over the 20 images simulated.
	\vspace{-0.1in}}
	\label{table:oracle}
\end{table}
\end{landscape}

\paragraph{A Generalized Formulation}
A binary quanta sensor artificially ``squeezes'' noise in the high-flux regime \cite{fossumModelingPerformanceSingleBit2013}, and therefore one may say that exposure-referred noise is a more appropriate objective than the binomial mean-squared error (MSE) considered so far.

To accommodate a more general analysis we define the following image-space loss function
\begin{equation}
    \mathcal{L}_{im.} \coloneqq \sum_i \frac{1}{W_i} E_i,
        \label{eq:optimal_allocation_loss_general}
\end{equation}
where $E_i$ is an arbitrary per-pixel normalized loss which is subsequently driven down by averaging $W_i$ measurements.
$E_i$ is therefore assumed to be a function of $H_i$ (or equivalently, $Y_i$).
We maintain the same constraint on the total number of detections:
\begin{equation}
    \sum_i W_i Y_i = D_T.
\end{equation}

Applying the method of Lagrange multipliers to the loss function $\mathcal{L} \coloneqq \mathcal{L}_{im.} + \lambda \left(\sum_i Y_i W_i - D_T\right)$, yields the optimal allocation
\begin{equation}
    W_i^{opt.} = D_T \left(
                        \frac{\sqrt{\nicefrac{E_i}{Y_i}}}
                            {\sum_j Y_j \sqrt{\nicefrac{E_j}{Y_j}}}
                    \right)
                \propto \sqrt{\frac{E_i}{Y_i}}.
\end{equation}
It can be verified that on setting $E_i = Y_i (1 - Y_i)$, the $\sqrt{1 - Y}$ weighting of Eq. \eqref{eqn:optimal-MSE} is recovered.

\begin{figure*}[t!]
    \centering
    \includegraphics[width=\linewidth]{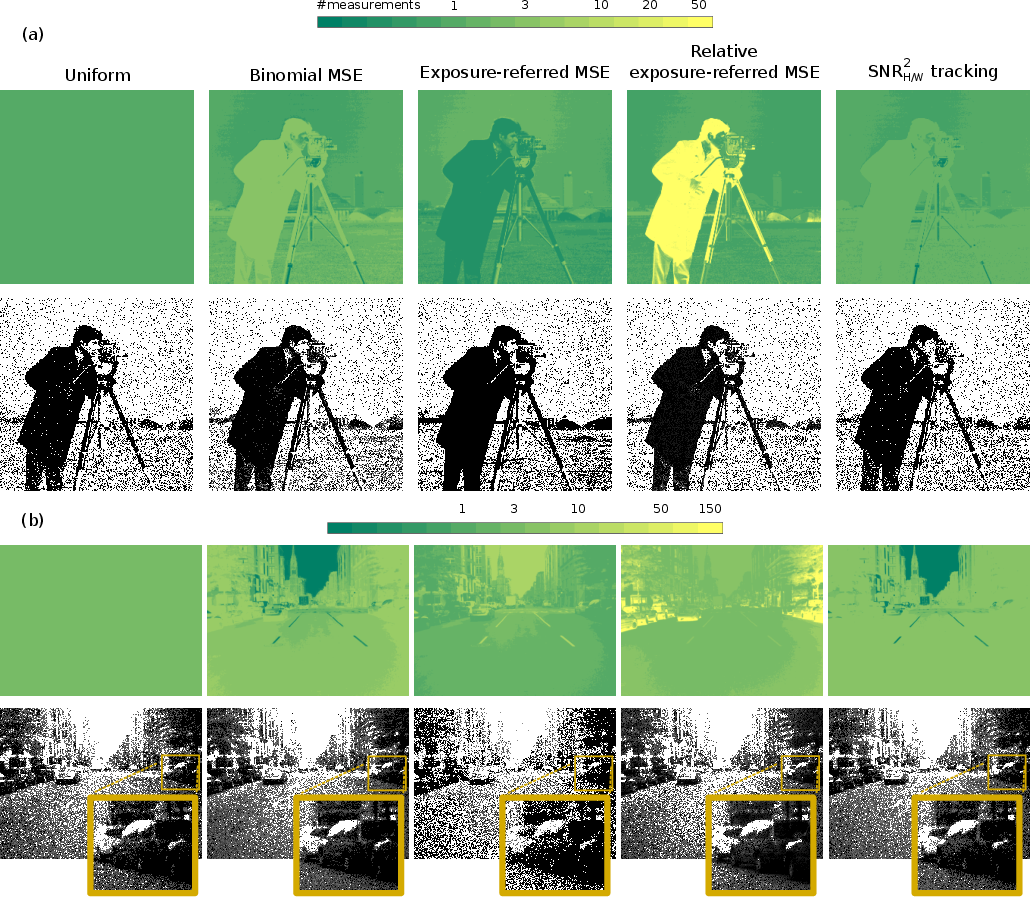}
    \caption{Two example images simulated under optimal allocations for the metrics of Table \ref{tab:optimal_measurement_allocation_DTonly}, in bottom rows of (a) and (b).
    The pseudo-color images in the top rows display the number of binary exposures allocated to each pixel by the expressions in the table (the colormap is in log-scale) -- the total expected number of detections is held constant for all allocations.
    The actual images are simulated by rounding those expressions, and assuming a mean flux of $1.59$ photons/pixel over the complete image for a single binary exposure (the peak of the $\snr_H$ curve \cite{chanWhatDoesOneBit2022}).
    Pixels with zero measurements are replaced with either zero or the maximum flux of the true image, as appropriate for each individual metric (the choice is independent of the image).
    The binomial MSE, the relative exposure-referred MSE, and the $\snr_{H/W}$-tracking loss function result in denser allocation towards dim pixels (and generally improved image quality), while the plain exposure-referred MSE does the opposite in both aspects.
    The peaky nature of the allocation with relative exposure-referred MSE can be seen through its strong highlights.\\
    (The reader is requested to zoom in to observe finer details.)}
    \label{fig:compare_meas_alloc_DTonly}
\end{figure*}

Suppl. Table \ref{tab:optimal_measurement_allocation_DTonly} shows three other possible loss functions with the optimal allocations obtained using the process above.
The binomial MSE has already been discussed.
``Exposure-referred MSE'' transfers the error to the linear radiance domain, for which the optimal allocation is proportional to $\frac{1}{\sqrt{1 - Y}}$.
This loss function has a potential problem: the optimal allocation \emph{diverges} for very high flux ($Y_i \to 1$).
Further, this metric generally encourages measurements to be spread more densely over \emph{bright} pixels -- the opposite of the binomial MSE discussed previously.
Images acquired under this allocation do have slightly more detail (less noise) in highlights compared to both a uniform spread as well as weighting by $\sqrt{1 - Y}$, but at the cost of almost complete loss of detail in dark regions (Suppl. Fig. \ref{fig:compare_meas_alloc_DTonly}).

The problem with exposure-referred MSE is partially compensated by defining a \emph{relative} version, which normalizes it by the squared radiance -- minimizing this is equivalent to maximizing $\snr_H$ directly.
The optimal allocation is now proportional to $\frac{1}{H \sqrt{1 - Y}}$.
Images acquired under this allocation do preserve much more detail in dim regions (Suppl. Fig. \ref{fig:compare_meas_alloc_DTonly}) -- perhaps excessively so, in fact, since the allocation now diverges for $H \to 0$ as well!
This is particularly a problem under high dynamic range: while we can reasonably assume a finite upper bound on flux, there is no obvious lower limit of flux.

The issues with exposure-referred error measures stem from the under-specified nature of the problem so far.
A simple approach to addressing the issue of diverging allocations is to regularize the problem: clamping the detection rate $Y$ to the range $[0.01, 0.99]$ can be considered one form of this.
Another principled alternative would be to formally include the several forces limiting the total number of measurements allowed: the amount of camera and scene motion, readout bandwidth and power consumption, and the latency-sensitivity of the vision task.
Placing explicit inequality constraints on total measurements results in a more complex optimization problem which does not have a closed-form solution. This is an important question which we do not explore further here and leave for future work.

The fourth row of Suppl. Table \ref{tab:optimal_measurement_allocation_DTonly} places the saturation look-ahead policy of the main paper in the current context.
In Fig. 4 of the main paper, the expected number of detections $\EX[D_i]^{opt.}$ tracks the $\snr_{H/W}$ plot in log-log space, which suggests a power-law relation (at least approximately).
Working backwards from that result to the loss function for which that allocation is optimal, provides insight into the behavior of the policy (which was definitely \emph{not} designed with any formal optimization problem in mind, only to yield adequate performance and be practically feasible -- see Suppl. Sec. \ref{sec:implementation_saturation_lookahead} for related discussion).
Focusing on the $k = 2$ case from the table, we see that the so-called ``loss'' function that the saturation look-ahead policy appears to minimize is a sum of (powers of) $\snr$s, which does seem extremely counter-intuitive.
A way to make sense of this observation is to realize that if treated abstractly, the function $\snr_{H/D}^2 \cdot \snr_{H/W}^2$ is larger for dim pixels, and therefore an allocation that favors dim pixels drives down the ``loss'' further.
This is the essence of the behavior we intuitively seek from an inhibition policy: $\sqrt{1 - Y}$ weighting does the same.
However, this is by no means an ideal metric, and tracking the $\snr_{H/W}$ curve precisely is not an absolute necessity.
Even confining ourselves to the space of the four choices considered in Suppl. Table \ref{tab:optimal_measurement_allocation_DTonly}, the allocation under binomial MSE appears to yield visually better images than the $\snr_{H/W}^k$-tracker, and so an inhibition policy that can practically realize it is likely to perform even better than saturation look-ahead.

\begin{landscape}
\begin{table}[h!]
    \centering
    \caption{Optimal measurement allocations and expected number of detections for four loss functions. $\snr_{H/D}$ is defined as $\sqrt{\snr_{H/D}^2} = \frac{H \sqrt{1 - Y}}{Y}$, and similarly $\snr_{H/W} \coloneqq \sqrt{\snr_{H/W}^2} = \frac{H \sqrt{1 - Y}}{\sqrt{Y}}$.}
    \label{tab:optimal_measurement_allocation_DTonly}
    \begin{tabular}{r||c|l|l}
    \toprule
        \textbf{Metric name}
            &   $E_i$
            &   $W_i^{opt.}$
            &   $\EX[D_i]^{opt.} = Y_i W_i^{opt.}$   \\
    \midrule
        Binomial MSE
            &   $Y_i (1 - Y_i)$
            &   $\propto \sqrt{1 - Y_i}$
            &   $\propto Y_i \sqrt{1 - Y_i}$  \\
    \midrule
        Exposure-referred MSE
            &   $\left(\frac{\mathrm{d}H_i}{\mathrm{d}Y_i}\right)^2
                \cdot Y_i (1 - Y_i)$
            &   $\propto \frac{1}{\sqrt{1 - Y_i}}$
            &   $\propto \frac{Y_i}{\sqrt{1 - Y_i}}$\\
        {}
            &   $= \frac{Y_i}{1 - Y_i}$
            &
            &   \\
    \midrule
        \emph{Relative} exposure-referred MSE
            &   $\frac{1}{H_i^2}
                \cdot \left(\frac{\mathrm{d}H_i}{\mathrm{d}Y_i}\right)^2
                \cdot Y_i (1 - Y_i)$
            &   $\propto \frac{1}{H_i \sqrt{1 - Y_i}}
                = \frac{\snr_{H_i/D_i}}{\snr_{H_i/W_i}^2}$
            &   $\propto \frac{Y_i}{H_i \sqrt{1 - Y_i}}
                = \frac{1}{\snr_{H_i/D_i}}$   \\
        ($\equiv \frac{1}{\snr_H^2}$)
            &   $\hspace{3em}   = \frac{1}{H_i^2} \frac{Y_i}{1 - Y_i}$
            &   $\hspace{4.5em} = \frac{1}{Y_i \snr_{H_i/D_i}}$
            &   \\
    \midrule
        \emph{$\snr_{H/W}$-tracker}
            &   $\frac{1}{Y_i} \snr_{H_i/W_i}^{2k}$
            &   $\propto \frac{1}{Y_i} \snr_{H_i/W_i}^k$
            &   $\propto \snr_{H_i/W_i}^k$   \\
        {}
            &   $k \in \{1, 2, \ldots\}$
            &   {}
            &   {}   \\
        For $k = 2$
            &   $\snr_{H_i/D_i}^{2} \cdot \snr_{H_i/W_i}^{2}$
            &   $\propto \frac{1}{Y_i} \snr_{H_i/W_i}^2
                    = \snr_{H_i/D_i}^2$
            &   $\propto \snr_{H_i/W_i}^2$   \\
    \bottomrule
    \end{tabular}
\end{table}
\end{landscape}

\paragraph{Is there a point to oracle-type analysis?}
These optimal allocations are derived assuming the pixel intensities are known ahead of time.
These oracle policies would not appear to be realistic, yet recent work on optimal spatially varying exposures proposes a two-step capture sequence with the first step being a pilot image that guides configuration \cite{quSpatiallyVaryingExposure2024}.
A similar pilot image approach may be useful for practical implementation of the optimal measurement allocations discussed here.

\subsection{Binary Rate Efficiency Metrics}
The efficiency metrics developed in Sec.~4.1 of the main manuscript evaluate the SNR of the measurement of photon exposure ($H$).
An alternative representation of the scene is the binary rate ($Y$).
Exposure and binary rate are related by a nonlinear transform of $H = -\ln(1-Y)$.
Entropy is a possible alternative to SNR for single-photon image sensors \cite{gnanasambandamExposureReferredSignaltoNoiseRatio2022} if the application processes binary rate images.
With a binary rate of $Y$ the entropy is
\begin{equation}
	S(Y) = -Y\log_2(Y) - (1-Y)\log_2(1-Y)
\end{equation}
with a maximum of 1 bit when $Y=0.5$ \cite{gnanasambandamExposureReferredSignaltoNoiseRatio2022}.
Similar to the main manuscript, an entropy detection efficiency may be defined as $S(Y)^2/Y$.
Like detection efficiency in the main manuscript, this metric also demonstrates that detections of nearly saturated pixels are less informative and should be inhibited for energy-efficient single-photon imaging.

\FloatBarrier

\section{Background on Static Single-Pixel Inhibition Policies \label{sec:static-single-pixel}}
Current single-photon sensor designs implicitly inhibit photons by setting a maximum count \cite{ogi250fps124dBDynamicRange2021, ota37W143dBDynamicRange1Mpixel2022} or lengthening the exposure time \cite{takatsuka36UmpitchSPAD2023} to limit detections and reduce avalanche energy.
These architectures comprise a family of inhibition policies that operate at the individual pixel level and do not adapt as a function of the history of photon detections.
Below is background on these already existing policies because our proposed inhibition policies build on top of these.

\renewcommand{\tabularxcolumn}[1]{>{\small}m{#1}}
\begin{figure}[b!]
	\centering
	\begin{tabularx}{0.7\columnwidth}{X}
		\hline
		\small
		\adjustbox{max width=0.8\linewidth}{
	\begin{tikzpicture}[baseline=(current bounding box.center)]
		\newcommand\phht{0.75}
		\newcommand{\upph}[2]{\draw[->, thick] (#1, 0) -- (#1, #2)}
		\newcommand{\upphin}[2]{\draw[->, thin] (#1, 0) -- (#1, #2) node[midway]{X}}

  		\node (L) at (-1.25, 2.1) {\large (a)};

		\node (A) at (-1, 1.5) {Inhibit};
        \node (Z) at (-1, 1.2) {Window};
		\node (B) at (-1, 0.5) {Photons};
		\node (O) at (0,0) {};
		\node (D) at (5, 0) {};
		\node (D2) at (7.9, -0.3) {time};
		
		\draw[->] (0,0) to (8,0);
		\upph{1}{\phht};
		\upphin{2}{\phht};
		
		\draw[draw=none, fill=black!15] (1,1) rectangle (2.5,1.75);
		\draw[<->] (1, 2) -- (2.5,2) node[midway,above] {$\tau_D$};
		
		\upph{4}{\phht};
		\draw[draw=none, fill=black!15] (4,1) rectangle (5.5,1.75);
		\draw[<->] (4, 2) -- (5.5,2) node[midway,above] {$\tau_D$};
		
		\upph{5.75}{\phht};
		\upphin{6}{\phht};
		\upphin{6.5}{\phht};
		\upphin{6.75}{\phht};
		
		\draw[draw=none, fill=black!15] (5.75,1) rectangle (7.25,1.75);
		\draw[<->] (5.75, 2) -- (7.25, 2) node[midway,above] {$\tau_D$};
	\end{tikzpicture}
 }   \\
		\hline
		\small
		\adjustbox{max width=0.8\linewidth}{
	\begin{tikzpicture}[baseline=(current bounding box.center)]
		\newcommand\phht{0.75}
		\newcommand{\upph}[2]{\draw[->, thick] (#1, 0) -- (#1, #2)}
		\newcommand{\upphin}[2]{\draw[->, thin] (#1, 0) -- (#1, #2) node[midway]{X}}
		\newcommand{\expline}[2]{\draw[-, thin, color=white] (#1, 1.75) -- (#1, 2.25); \draw[<->] (#1, 2) -- (#1+#2, 2) node[midway,above] {$T$}}

  		\node (L) at (-1.25, 2.1) {\large (b)};

		\node (A) at (-1, 1.5) {Inhibit};
        \node (Z) at (-1, 1.2) {Window};
		\node (B) at (-1, 0.5) {Photons};
		\node (O) at (0,0) {};
		\node (D) at (5, 0) {};
		\node (D2) at (7.9, -0.3) {time};
		
		\draw[->] (0,0) to (8,0);
		\expline{0}{2};
		\expline{2}{2};
		\expline{4}{2};
		\expline{6}{2};
		
		\upph{1}{\phht};
		\draw[draw=none, fill=black!15] (1,1) rectangle (2,1.75);
		\upphin{1.5}{\phht};
		
		\upph{3.5}{\phht};
		\draw[draw=none, fill=black!15] (3.5,1) rectangle (4,1.75);
		
		\upph{6.25}{\phht};
		\draw[draw=none, fill=black!15] (6.25,1) rectangle (8,1.75);
		\upphin{7}{\phht};
		\upphin{7.5}{\phht};
		
	\end{tikzpicture}
 }   \\
		\hline
		\small
		\adjustbox{max width=1\linewidth}{
	\begin{tikzpicture}[baseline=(current bounding box.center)]
		\newcommand\phht{0.75}
		\newcommand{\upph}[2]{\draw[->, thick] (#1, 0) -- (#1, #2)}
		\newcommand{\upphin}[2]{\draw[->, thin] (#1, 0) -- (#1, #2) node[midway]{X}}
		\newcommand{\expline}[3]{\draw[-, thin, color=white] (#1, 1.75) -- (#1, 2.25); \draw[<->] (#1, 2) -- (#1+#2, 2) node[midway,above] {#3}}

  		\node (L) at (-1.25, 2.1) {\large (c)};

		\node (A) at (-1, 1.5) {Inhibit};
        \node (Z) at (-1, 1.2) {Window};
		\node (B) at (-1, 0.5) {Photons};
		\node (O) at (0,0) {};
		\node (D) at (5, 0) {};
		\node (D2) at (9.5, -0.3) {time};
		
		\draw[->] (0,0) to (9.5,0);
		\expline{0}{0.5}{$T_1$};
		\expline{0.5}{0.5}{$T_1$};
		\expline{1}{0.5}{$T_1$};
		\expline{1.5}{1.5}{$T_2$};
		\expline{3}{1.5}{$T_2$};
		\expline{4.5}{2.5}{$T_3$};
		\expline{7}{2.5}{$T_3$};
		
		\upph{0.25}{\phht};
		\draw[draw=none, fill=black!15] (0.25,1) rectangle (0.5,1.75);
		
		\upph{1.125}{\phht};
		\draw[draw=none, fill=black!15] (1.125,1) rectangle (1.5,1.75);
		
		\upph{2}{\phht};
		\draw[draw=none, fill=black!15] (2,1) rectangle (3,1.75);
		\upphin{2.75}{\phht};
		
		\upph{3.75}{\phht};
		\draw[draw=none, fill=black!15] (3.75,1) rectangle (4.5,1.75);
		
		\upph{4.75}{\phht};
		\draw[draw=none, fill=black!15] (4.75,1) rectangle (7,1.75);
		\upphin{5.5}{\phht};
		\upphin{6}{\phht};
		
		\upph{8.25}{\phht};
		\draw[draw=none, fill=black!15] (8.25,1) rectangle (9.5,1.75);
		\upphin{8.5}{\phht};
		\upphin{9}{\phht};
		
	\end{tikzpicture}
 }   \\
		\hline
	\end{tabularx}
	\caption{\textbf{Single-pixel static inhibition policies without computations.}
		Arrows represent incoming photons with an `X' for inhibition.
		(a) \textit{Asynchronous recharge with dead time ($\tau_D$)}.
		After a photon detection, the bias voltage of a SPAD must be recharged.
		During this dead time ($\tau_D$) photons are inhibited.
		(b) \textit{Clocked recharge}.
		The recharge period of $T$ sets a window in which 0 or 1 photons can be detected.
		Subsequent photons within the window are inhibited.
		(c) \textit{Clocked recharge with exposure brackets}.
		An extension of clocked recharge with a sequence of different periods.}
	\label{table:static_policies}
	\vspace{-0.15in}
\end{figure}
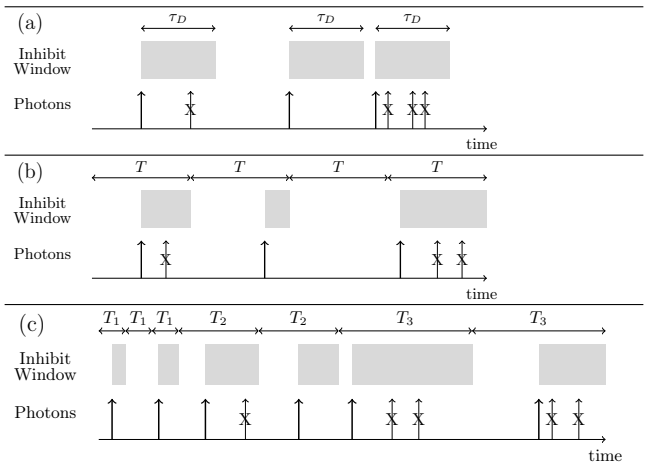

Suppl. Fig.~\ref{table:static_policies} shows a subset of static single-pixel inhibition policies.
SPADs require recharge after an avalanche-inducing photon detection during which recording a subsequent photon is not possible (dead time, $\tau_D$).
This detector response of \textit{asynchronous recharge with dead time} inhibits photons at high exposure \cite{henderson25625640nm2019, ingleHighFluxPassive2019}, yet, power consumption is excessive when the average inter-photon arrival interval is shorter than the SPAD dead time \cite{ota37W143dBDynamicRange1Mpixel2022, morimotoMegapixelTimegatedSPAD2020}.
\textit{Clocked recharge} is an alternative that establishes time windows, similar to a conventional exposure time, during which 0 or 1 photon may be detected \cite{takatsuka36UmpitchSPAD2023, ota37W143dBDynamicRange1Mpixel2022}.
After the first photon, any subsequent arrivals during the same predefined exposure window are inhibited.
The average number of inhibited photons is equal to $\sum_{k=2}^{\infty} (k-1) \prob(K=k ; H)$
where $\prob$ denotes the Poisson probability mass function.
At the measurement-limited SNR-optimal exposure $H = 1.6$ there is a $Y=0.80$ chance of detecting a photon with an average of $0.83$ photons inhibited per measurement. With $H{\gg}1$ the average number of inhibited photons approaches $H-1$ yet the signal-to-noise ratio degrades because the pixel is nearly saturated.
Clocked recharge considerably reduces power in bright light as compared to asynchronous recharge \cite{morimotoMegapixelTimegatedSPAD2020, takatsuka36UmpitchSPAD2023}.
Because of this, our proposed policies typically maintain and extend clocked recharge.

\begin{figure}[t!]
	\centering
	\includegraphics[width=0.65\textwidth]{{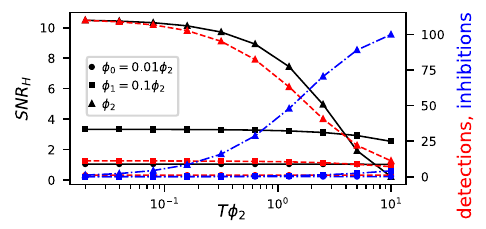}}
	\vspace{-0.2in}\caption{(left) The $SNR_H$ (black, $-$) versus the exposure time with the total sensing latency, $T_L$, maintained by varying the number of measurements ($W$ = $T_L/T$) at three different flux levels. (right) The number of detections (red, $-{}-$) and the number of inhibitions (blue, {$- \cdot$}$-$). A vertical slice represents one exposure time of an exposure bracket policy.}
	\label{fig:snr_vs_texp} 
	\vspace{-0.15in}
\end{figure}

\textit{Clocked recharge with exposure brackets} \cite{duttonHighDynamicRange2018, takatsuka36UmpitchSPAD2023}, shown in Suppl. \cref{table:static_policies}(c), is a static inhibition policy that uses multiple exposure times to balance constraints on detections and measurements while maintaining SNR over a range of illumination levels.
Longer exposure times measure dim pixels with good SNR and limit the detections of bright pixels; short exposure times measure bright pixels with good SNR.
Suppl. Fig.~\ref{fig:snr_vs_texp} guides the tradeoffs between detections, inhibitions, and SNR when selecting a single exposure time of a bracketing sequence.
Suppl. Table~ \ref{table:exposure_brackets} selects three specific exposure times of an exposure bracket inhibition policy and tracks detections, inhibitions, and the contributions of each exposure time to the HDR reconstruction.
Due to near saturation, the detections by the brightest pixel at the longest exposure time(s) have a low weighting for SNR-based HDR reconstruction \cite{gnanasambandamHDRImagingQuanta2020} but still represent $10/25.7 = 38.9\%$ of the total detections, suggesting a clear opportunity for more advanced inhibition policies to reduce avalanche power.

\begin{table}[!htbp]
\centering
\begin{tabularx}{0.68\columnwidth}{l|lll|lll|lll|l}
\toprule
{} & \multicolumn{3}{c|}{$T=0.1/\phi_2$} & \multicolumn{3}{c|}{$T=1.0/\phi_2$} & \multicolumn{3}{c|}{$T=10.0/\phi_2$} & HDR \\
$\phi [\phi_2]$ &  wt. &  D & I & wt. & D & I & wt. &  D & I & SNR \\
\midrule
0.01 &   0.10 &  0.01 & 0.00 & 0.09 & 0.10 & 0.00 & 0.90 &  0.95 &  0.05 &     1.03              \\
0.10 &   0.32 &  0.10 & 0.00 & 0.14 & 0.95 & 0.05 & 0.85 &  6.32 &  3.68 &     2.62            \\
1.00 &   0.98 &  0.95 & 0.05 & 0.85 & 6.32 & 3.68 & 0.01 & \textbf{10.00} & 88.11 &     2.61  \\
\midrule
Total & {} &  1.06 & 0.05 &  {} & 7.37 & 3.73 & {} &  17.27  & 91.83 & {}     \\
\bottomrule
\end{tabularx}
	\caption{Clocked recharge with exposure bracket results for three pixel fluxes with each exposure time using $W=10$ of measurements.
		Flux values are in units of the maximum flux $\phi_2$.
		\textit{wt.} is the weighting for HDR reconstruction \cite{gnanasambandamHDRImagingQuanta2020}, \textit{D} is detections, and \textit{I} is inhibitions.
		A bold value indicates an opportunity to improve detection efficiency by using a more advanced inhibition policy.}
	\label{table:exposure_brackets}
	\vspace{-0.1in}
\end{table}
\FloatBarrier

\section{Details of Experiments on Dynamic Scenes}
This section details the methods for Section 7 of the main text and motivates the sub-sampling ratio chosen for comparison in Figure 7.

\subsection{Sub-sampling Factor Tradeoff}
\begin{figure}
	\includegraphics[width=\linewidth]{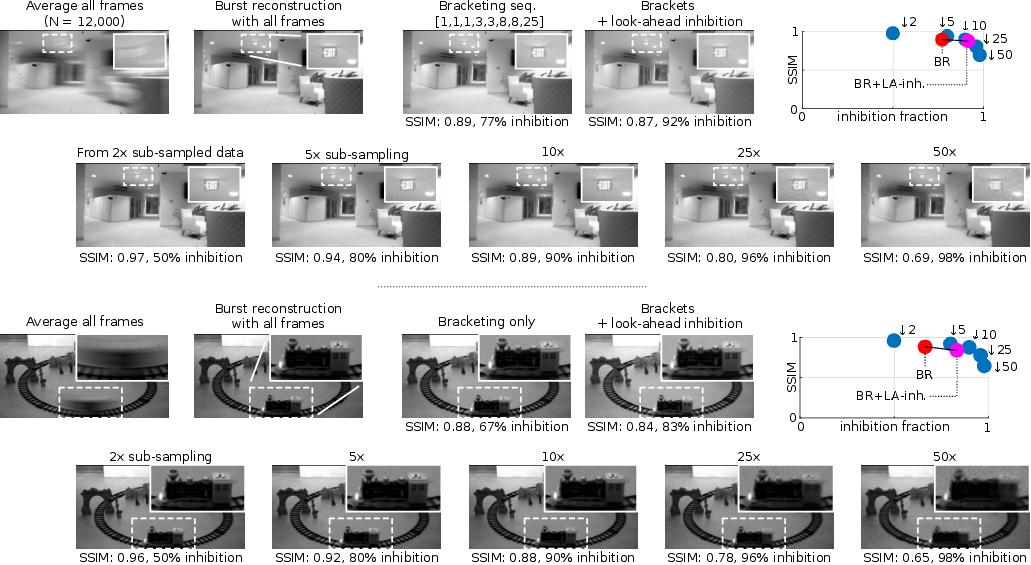}
	\vspace{-0.2in}
	\caption{\textbf{Quality versus inhibition for a single image}.
		For two separate sequences of $12{,}000$ binary frames each, burst reconstruction \cite{maQuantaBurstPhotography2020a} is performed first directly (with all photons), and then with various inhibition policies:
		(top) exposure bracketing, with and without saturation look-ahead inhibition, and
		(bottom) sub-sampling by dropping frames.
		The binary frames were captured using the SwissSPAD2 sensor \cite{ulku512512SPAD2019}, similar to the main paper.
		The plots on the top-right measure the image quality relative to the reference (no-inhibition) result using SSIM \cite{wang2004image}, versus the fraction of photons inhibited/dropped by the policy.
		Sub-sampling factors larger than $10\times$ incur substantial image quality loss for these two images: analogous results are expected for other scenes, possibly for different sub-sampling factors depending on light levels.
		Separately, the saturation look-ahead policy provides significant inhibition on top of bracketing, with minimal loss in image quality.}
	\label{fig:qbp_single_image}
	\vspace{-0.1in}
\end{figure}

	Sub-sampling a binary frame sequence is equivalent to setting a longer period for clocked recharge, but with the distinction that the actual exposure duration for which the pixel is photo-sensitive is kept constant (the length of the original binary frame exposures).
	Suppl. Fig. \ref{fig:qbp_single_image} shows example results for two real images with varying sub-sampling factors, evaluated on image quality using SSIM \cite{wang2004image}.
	For sub-sampling factors greater than $10\times$ a large drop in SSIM can be seen, and similar behavior was obtained with the images of Fig. 7 in the main paper.

	\subsection{Additional Results}
		A video of the entire sequence of Figure 7 of the main text is viewable at the project webpage: \url{https://wisionlab.com/project/inhibition}.

\newpage
\section{Implementation of Proposed Policies}
In Sec. \ref{sec:calculation-based-implementation} we \textit{estimate} the circuitry required to implement the proposed calculation-based inhibition policy (of Fig. 2).
A design assumption is that the implementation will be more constrained by area than computation latency --- at $\SI{400}{MHz}$ logic clock frequency \cite{ardeleanComputationalImagingSPAD} 4,000 clock cycles are available for computation during a $\SI{10}{\mu s}$ recharge period.
Therefore, the approach prioritizes minimization of the required in-pixel memory.
We emphasize that these are ``back of the envelope'' estimates; we have not fabricated a chip or created synthesizable digital logic yet.
Sec. \ref{sec:implementation_saturation_lookahead} describes the bracketing-based saturation look-ahead policy in terms of the computation and memory required, as well as the likelihood-maximization process used to convert the bracketed measurements to an estimate of incident flux.

\subsection{Calculation-based Inhibition}\label{sec:calculation-based-implementation}
	As a reminder from the main text, the inhibition score at each pixel is calculated as
	\begin{equation}
		S(i,j,t) = K \ast [(2F(i,j,t)-\mathbf{1})\cdot M(i,j,t)] \label{eq:inhibition_score}
	\end{equation}
	which applies a spatio-temporal filtering kernel, $K$, of dimensions $L,H,T$ to a ternary representation of the pixel result ($1$, $0$, or $-1$ for a detection, a disabled pixel, or a measurement that does not detect a photon, respectively).
	The kernel $K$ can typically be separated into spatial and temporal components as $K = K_s \otimes K_t$ with dimensions $L \times H \times 1$ and $1 \times 1 \times T$, respectively.
	After each binary frame, the score is compared to a threshold $\eta$ and the pixel is disabled for the subsequent $\tau_H$ frames:
	$M(i,j,t') = 0$ for $\{t' | t+1 \leq t' \leq t+1+\tau_H\}$ if $S(i,j,t) > \eta$.
	Suppl.~Sec.~\ref{sec:imaging_policies} shows the spatial and temporal kernels ($K_s$ and $K_T$) used in the simulations.

	Suppl. Table~\ref{table:score_implementation} describes possible on-chip and in-pixel circuitry for the calculation-based inhibition policy.
	A subset of the circuit elements must be independent for each pixel as indicated by an entry of ``no'' in the column titled ``Share?''.
	However, other computation circuitry could be shared among a local neighborhood of pixels and housed in a macropixel computation unit \cite{ardeleanComputationalImagingSPAD, henderson25625640nm2019}. The circuitry is separated into subcircuits of:

	\begin{enumerate}
		\item \textit{SPAD control}: directly controls the bias voltage of the SPAD and allows for enable/disable of the pixel.
		\item \textit{Inhibition score}: short-length memory for detection results and arithmetic circuits (adder and shift left) for spatio-temporal computations.
		\item \textit{Inhibition control}: evaluates if the score, $S$, exceeds the inhibition threshold, $\eta$. If $S>\eta$, disables the pixel for a count of $\tau_H$ clocked-recharge exposure periods.
		\item \textit{Measurement results}: in-pixel counters to record the number of detections and inhibitions. These results must be readout to reconstruct the image.
	\end{enumerate}
	\begin{landscape}
	\setlength{\tabcolsep}{12pt}
	\begin{longtable}{  @{}>{\raggedright\arraybackslash} p{0.1\linewidth}    p{0.27\linewidth}   >{\raggedright\arraybackslash}p{0.1\linewidth}    p{0.27\linewidth}    p{0.08\linewidth} @{} }
		\caption{Circuitry to implement the calculation-based inhibition policy.
			The SPAD control and detection counter is part of the circuitry of a conventional SPAD pixel (for example, see Ota et al. \cite{takatsuka36MuMpitch2023}).
			The last column indicates whether computations may be shared (time-multiplexed) among multiple pixels at a macropixel arithmetic unit.
			The order of implementation to store the single-pixel results of the temporal filter ($K_T$) and then re-calculate $K_s$ at the end of each exposure period is designed to to minimize the required in-pixel memory (since the results of $K_s$ span a wider range than $K_T$).
			The measurement results subcircuit is needed to recreate the total photon rate after readout.
			This requires a count of the detections and a count of the number of exposures during which the pixel was active or disabled.
			An approach that reduces the required circuitry counts only the inhibitions starts with the total number of disabled frame periods as $starts \times \tau_H$.}
		\label{table:score_implementation} \\
		\toprule
		\textbf{Circuit   element} & \textbf{Description}                                                                                                       & \textbf{Subcircuit} & \textbf{Notes}                                                                                                                                                                                                & \textbf{Share?} \\
		\midrule
		PMOS transistor            & SPAD bias control                                                                                                          & SPAD control        &                                                                                                                                                                                                               & no                                 \\
		\midrule
		OR gate                    & Logic to determine SPAD enable, and drive PMOS gate                                                                         & SPAD control        &                                                                                                                                                                                                               & no                                 \\
		\midrule
		Logic gate                    & Converts SPAD detection and SPAD enable state to two bit ternary of 1, 0, -1.                                                               & Inhibition score    &                                                                                                                                                                                                               & no                                 \\
		\midrule
		Register{[}1:0{]} $\times 4$       & 2-bit register to store signed detection results.  $T$ copies arranged as a shift register to   create the temporal filter.       & Inhibition score    & Implements $K_T$.  Proposed has $T=4 \Rightarrow \; 8$~ registers as $K_T=[1, 1, 1, 1]$.                                                                                                                                              & no                                 \\
		\midrule
		Adder                      & Adds each of T detection results.                                                                                           & Inhibition score    & Implements $K_T$. Proposed allows results from [-4, 4].                                                                                                                                                     & yes                                \\
		\midrule
		Shift left                 & $K_s$ multiplications that are constrained to powers to 2.                                                                  & Inhibition score    & Implements $K_s$. Must allow for $\times(-1)$. Proposed needs max shift of $\ll 3$  (for $\times 8$).                                                                                                             & yes                                \\
		\midrule
		Adder                      & Adds pixel and nearest neighbor pixels for a spatial convolution.                                                                                    & Inhibition score    & Implements $K_s$. Proposed results from {[}-64, 64{]}. This result does not need to be stored.                                                                                                              & yes                                \\
		\midrule
		Comparison                 & $S > \eta$                                                                                         & Inhibition control    & Proposed implementation has common $\eta$ among all pixels.                                                                                                                                      & yes                                \\
		\midrule
		Counter                    & Count for holdoff period, $\tau_H$, triggered by $S>\eta$. Counts recharge periods and releases SPAD disable. & Inhibition control  & Output combines with global clock recharge to enable/disable SPAD. The proposed parameters in simulations require a maximum counter depth of 5.                                                                                                                                            & no                                 \\
		\midrule
		Counter                    & Count of detections. Depth set by maximum number of frames.                                                            & Measurement results & Readout to create the final pixel intensity value. Simulations used a maximum of 1,000 frames which would require a 10 bit counter.                                                                                                                                                               & no                                 \\
		\midrule
		Counter                    & Count of inhibition starts. Depth less than detection counter by   $\times \log_2(\tau_H)$. For the best performing imaging policy $\tau_H=32$ such that a 5 bit counter would be required to count inhibition starts ($2^{10}/2^5 = 2^5$).                             & Measurement results & Only counting inhibition \textit{starts} reduces the required depth of the counter. Readout to recreate   final pixel intensity value. End effects will be observed if the composite   frame ends during a holdoff time. & no  \\
		\bottomrule
	\end{longtable}
	\end{landscape}

\clearpage

\subsection{Saturation Look-ahead Inhibition \label{sec:implementation_saturation_lookahead}}
	This policy is described with an example in Fig. 4a of the main paper.

	\vspace{-1em}
	\paragraph{Pseudo-code}
		A MATLAB-like code listing describing the complete implementation is provided below.

\begin{lstlisting}[language=Octave]
function [B_LA, M] = bracket_LA(B, seq, policy)
% Look-ahead inhibition with exposure brackets:
% Input arguments:
%  B:   Nx1 binary vector (for single SPAD pixel)
%  seq: Tx1 bracketing sequence (integer) lengths
%       assumed sorted in non-decreasing order
%  policy: (length(unique(seq))-1) x 1, integer
% Outputs:
%  B_LA: Tx1 binary vector
%  M:    Nx1 "inhibition pattern" of line 299
%        M[n] = 1 means the pixel is _enabled_.
    B_LA = false([T 1]);
    M = true([N 1]);
    n = 1; sp = 0; Bsum_sp = 0; sat = false;
    for s = 1:length(seq)
        % Bracketing from binary SPAD frames
        for ns = 1:seq(s)
            B_LA(s) = B_LA(s) || (M(n) && B(n));
            if ns < seq(s)  % inh. within bracket
                M(n+1) = M(n) && ~B_LA(s);
            n = n + 1;
        % Look-ahead inhibition implementation
        if (s == 1) || (seq(s) ~= seq(s-1))
            % new unique sequence length
            sp = sp + 1; Bsum_sp = 0;            
        Bsum_sp = Bsum_sp + B_LA(s);
        if sat || (Bsum_sp > policy(sp))
            M(n) = 0; sat = true;   % disable
            
\end{lstlisting}

	\vspace{-1em}
	\paragraph{Complexity}
		The look-ahead inhibition policy is single-pixel and thus is expected to allow for a lighter-weight in-pixel implementation than the calculation-based policies.
		For a bracketing exposure time sequence of $T := \{T_i\}_{i=1}^{K}$, the measurement results are represented by the binary sequence of detections $B_T := \{b_i\}_{i=1}^{K}$ and the inhibition pattern $M_T := \{m_i\}_{i=1}^{K}$, where $m_i = 1$ denotes the pixel being enabled during that exposure.
		The memory footprint of both $B_T$ and $M_T$ is already relatively small, but further efficiency is realized by recognizing that for a given unique exposure time, the order of detections with that setting is not important; instead just the sum of detections contributes to the flux estimate.

		Furthermore, certain detection sequences are precluded by the inhibition policy.
		As an example, for the Fibonacci bracketing sequence $[1, 1, 2, 3, 5, 8, 13, 21]$ used in the main text with the inhibition policy of $[2, 1, 1, 1, 1, 1]$, only 15 unique combinations of $(B_T, M_T)$ are possible.
		Thus, inhibition may even result in \emph{greater} efficiency than the standard exposures, not less (at least in terms of memory and bandwidth use) --- the original Fibonacci brackets have $192$ possible unique measurements.
		When multiple bracketing cycles are aggregated on-chip, it may be possible to map the binary detection sequence to an index in a histogram via an encoding implemented on-chip, with only the histogram read out later.

	\vspace{-1em}
	\paragraph{Control signal flow.}
		At the conclusion of each exposure time sequence within a bracketing sequence, the number of detections must be compared to the threshold count in the inhibition policy.
		If the comparison triggers inhibition, this status is stored and used to disable the SPAD, as in the implementation of \ref{sec:calculation-based-implementation}.
		At the end of a bracketing sequence a global signal is required to reset the inhibition status of a pixel.

	\subsubsection*{Maximum Likelihood Estimation (MLE) Computation for Exposure Bracketing}
	As above, the bracketing exposure time sequence is denoted by $T := \{T_i\}_{i=1}^{K}$, and the inhibition pattern by $M_T := \{m_i\}_{i=1}^{K}$.
After bracketing, every sequence of $\mathrm{sum}(T) = \sum_i T_i$ binary measurements at the original rate is replaced with a binary sequence $B_T := \{b_i\}_{i=1}^{K}$.

The likelihood $L$ is given as a function of incident flux $\phi$:
\begin{equation}
    L(\phi) = \prod_{i=1}^{K}
                    \left[
                        \left(1 - m_i\right)
                        +
                        m_i
                        \cdot
                        \left(\exp(-\phi T_i)^{1 - b_i}
                                \cdot \left(1 - \exp(-\phi T_i)\right)^{b_i}
                        \right)
                    \right].
\end{equation}
This expression does not have a closed-form expression for its maximum in $\phi$.
Therefore, we optimize $\phi$ numerically given a particular combination of $M_T$ and $B_T$, searching exhaustively over $2{,}000$ uniformly spaced points in the range $[0, 10]$.
For a fixed bracket cycle $T$, the MLE may be found offline for all possible combinations of $B_T$ and $M_T$ and stored in a look-up table (LUT) of maximal possible size $2^{2 \times \mathrm{count}(T)}$.
But as stated above, only 15 sequences of detections are possible for the Fibonacci bracketing sequence used in the main text when combined with saturation look-ahead inhibition.
Therefore the corresponding LUT is also extremely small in practice.

{
	\small
	\bibliographystyle{ieeenat_fullname}
	\bibliography{reflist}

\begin{thebibliography}{10}
\providecommand{\url}[1]{\texttt{#1}}
\providecommand{\urlprefix}{URL }
\providecommand{\doi}[1]{https://doi.org/#1}

\bibitem{arbelaezContourDetectionHierarchical2011}
Arbel{\'a}ez, P., Maire, M., Fowlkes, C., Malik, J.: Contour {{Detection}} and
  {{Hierarchical Image Segmentation}}. IEEE Transactions on Pattern Analysis
  and Machine Intelligence  \textbf{33}(5),  898--916 (May 2011).
  \doi{10.1109/TPAMI.2010.161}

\bibitem{ardeleanComputationalImagingSPAD}
Ardelean, A.: Computational {{Imaging SPAD Cameras}}. Ph.D. thesis, EPFL
  (2023). \doi{10.5075/epfl-thesis-9501}

\bibitem{barlow1953summation}
Barlow, H.B.: Summation and inhibition in the frog's retina. The Journal of
  physiology  \textbf{119}(1), ~69 (1953)

\bibitem{berkovichScalable20202015}
Berkovich, A., {Datta-Chaudhuri}, T., Abshire, P.: A scalable 20 \texttimes 20
  fully asynchronous {{SPAD-based}} imaging sensor with {{AER}} readout. In:
  2015 {{IEEE International Symposium}} on {{Circuits}} and {{Systems}}
  ({{ISCAS}}). pp. 1110--1113 (May 2015). \doi{10.1109/ISCAS.2015.7168832}

\bibitem{bianHighresolutionSinglephotonImaging2023}
Bian, L., Song, H., Peng, L., Chang, X., Yang, X., Horstmeyer, R., Ye, L., Zhu,
  C., Qin, T., Zheng, D., Zhang, J.: High-resolution single-photon imaging with
  physics-informed deep learning. Nature Communications  \textbf{14}(1), ~5902
  (Sep 2023). \doi{10.1038/s41467-023-41597-9}

\bibitem{boso2015low}
Boso, G., Buttafava, M., Villa, F., Tosi, A.: Low-cost and compact
  single-photon counter based on a {CMOS} {SPAD} smart pixel. IEEE Photonics
  Technology Letters  \textbf{27}(23),  2504--2507 (2015)

\bibitem{canonspad}
{Canon Inc.}: {Canon Launches MS-500 - The World's First Ultra-High-Sensitivity
  Interchangeable-Lens SPAD Sensor Camera}.
  \url{https://www.usa.canon.com/newsroom/2023/20230801-ms500}, {Canon Press
  Release 8/1/2023. Accessed 2/25/2024.}

\bibitem{carey100000Fps2013a}
Carey, S.J., Lopich, A., Barr, D.R., Wang, B., Dudek, P.: A 100,000 fps vision
  sensor with embedded {{535 GOPS}}/{{W}} 256\texttimes 256 {{SIMD}} processor
  array. In: 2013 {{Symposium}} on {{VLSI Circuits}}. pp. C182--C183 (Jun 2013)

\bibitem{chanWhatDoesOneBit2022}
Chan, S.H.: What {{Does}} a {{One-Bit Quanta Image Sensor Offer}}? IEEE
  Transactions on Computational Imaging  \textbf{8},  770--783 (2022).
  \doi{10.1109/TCI.2022.3202012}

\bibitem{charbon3DStackedCMOSSPAD2018}
Charbon, E., Bruschini, C., Lee, M.J.: {{3D-Stacked CMOS SPAD Image Sensors}}:
  {{Technology}} and {{Applications}}. In: 2018 25th {{IEEE International
  Conference}} on {{Electronics}}, {{Circuits}} and {{Systems}} ({{ICECS}}).
  pp.~1--4 (Dec 2018). \doi{10.1109/ICECS.2018.8617983}

\bibitem{chi2020dynamic}
Chi, Y., Gnanasambandam, A., Koltun, V., Chan, S.H.: Dynamic low-light imaging
  with quanta image sensors. In: Computer Vision--ECCV 2020: 16th European
  Conference, Glasgow, UK, August 23--28, 2020, Proceedings, Part XXI 16. pp.
  122--138. Springer (2020)

\bibitem{della2020128}
Della~Rocca, F.M., Mai, H., Hutchings, S.W., Al~Abbas, T., Buckbee, K.,
  Tsiamis, A., Lomax, P., Gyongy, I., Dutton, N.A., Henderson, R.K.: A
  128$\times$128 {SPAD} motion-triggered time-of-flight image sensor with
  in-pixel histogram and column-parallel vision processor. IEEE Journal of
  Solid-State Circuits  \textbf{55}(7),  1762--1775 (2020).
  \doi{10.1109/JSSC.2020.2993722}

\bibitem{doi:10.1146/annurev-vision-102016-061345}
Diamond, J.S.: Inhibitory interneurons in the retina: {{Types}}, circuitry, and
  function. Annual Review of Vision Science  \textbf{3}(1),  1--24 (2017).
  \doi{10.1146/annurev-vision-102016-061345}

\bibitem{dollarStructuredForestsFast2013}
Dollar, P., Zitnick, C.L.: Structured {{Forests}} for {{Fast Edge Detection}}.
  In: Proceedings of the {{IEEE International Conference}} on {{Computer
  Vision}}. pp. 1841--1848 (2013)

\bibitem{duttonHighDynamicRange2018}
Dutton, N.A.W., Al~Abbas, T., Gyongy, I., Mattioli Della~Rocca, F., Henderson,
  R.K.: High {{Dynamic Range Imaging}} at the {{Quantum Limit}} with {{Single
  Photon Avalanche Diode-Based Image Sensors}}. Sensors  \textbf{18}(4), ~1166
  (Apr 2018). \doi{10.3390/s18041166}

\bibitem{fossumModelingPerformanceSingleBit2013}
Fossum, E.R.: Modeling the {{Performance}} of {{Single-Bit}} and {{Multi-Bit
  Quanta Image Sensors}}. IEEE Journal of the Electron Devices Society
  \textbf{1}(9),  166--174 (Sep 2013). \doi{10.1109/JEDS.2013.2284054}

\bibitem{frankeInhibitionDecorrelatesVisual2017}
Franke, K., Berens, P., Schubert, T., Bethge, M., Euler, T., Baden, T.:
  Inhibition decorrelates visual feature representations in the inner retina.
  Nature  \textbf{542}(7642),  439--444 (Feb 2017). \doi{10.1038/nature21394}

\bibitem{gabrielliFastReadoutPixel2008}
Gabrielli, A.: Fast readout for pixel devices. Measurement Science and
  Technology  \textbf{19}(8),  085101 (Jun 2008).
  \doi{10.1088/0957-0233/19/8/085101}

\bibitem{gallego2020event}
Gallego, G., Delbr{\"u}ck, T., Orchard, G., Bartolozzi, C., Taba, B., Censi,
  A., Leutenegger, S., Davison, A.J., Conradt, J., Daniilidis, K., et~al.:
  Event-based vision: A survey. IEEE Transactions on Pattern Analysis and
  Machine Intelligence  \textbf{44}(1),  154--180 (2020)

\bibitem{gardnerLearningPredictIndoor2017a}
Gardner, M.A., Sunkavalli, K., Yumer, E., Shen, X., Gambaretto, E., Gagn{\'e},
  C., Lalonde, J.F.: Learning to predict indoor illumination from a single
  image. ACM Trans. Graph.  \textbf{36}(6),  176:1--176:14 (Nov 2017).
  \doi{10.1145/3130800.3130891}

\bibitem{gnanasambandamHDRImagingQuanta2020}
Gnanasambandam, A., Chan, S.H.: {{HDR Imaging With Quanta Image Sensors}}:
  {{Theoretical Limits}} and {{Optimal Reconstruction}}. IEEE Transactions on
  Computational Imaging  \textbf{6},  1571--1585 (2020).
  \doi{10.1109/TCI.2020.3041093}

\bibitem{gnanasambandamExposureReferredSignaltoNoiseRatio2022}
Gnanasambandam, A., Chan, S.H.: Exposure-{{Referred Signal-to-Noise Ratio}} for
  {{Digital Image Sensors}}. IEEE Transactions on Computational Imaging
  \textbf{8},  561--575 (Jun 2022). \doi{10.1109/TCI.2022.3187657}

\bibitem{guptaFibonacci2013}
Gupta, M., Iso, D., Nayar, S.K.: Fibonacci {{Exposure Bracketing}} for {{High
  Dynamic Range Imaging}}. In: 2013 {{IEEE International Conference}} on
  {{Computer Vision}}. pp. 1473--1480. {IEEE}, {Sydney, Australia} (Dec 2013).
  \doi{10.1109/ICCV.2013.186}

\bibitem{gyongy2021direct}
Gyongy, I., Dutton, N.A., Henderson, R.K.: Direct time-of-flight single-photon
  imaging. IEEE Transactions on Electron Devices  \textbf{69}(6),  2794--2805
  (2021)

\bibitem{gyongy2021high}
Gyongy, I., Mart{\'\i}n, G.M., Turpin, A., Ruget, A., Halimi, A., Henderson,
  R., Leach, J.: High-speed vision with a {3D}-stacked {SPAD} image sensor. In:
  Advanced Photon Counting Techniques XV. vol. 11721, p. 1172105. SPIE (2021)

\bibitem{haldaneMETHODESTIMATINGFREQUENCIES1945}
Haldane, J.B.S.: {{On a Method Of Estimating Frequencies}}. Biometrika
  \textbf{33}(3),  222--225 (Nov 1945). \doi{10.1093/biomet/33.3.222}

\bibitem{henderson25625640nm2019}
Henderson, R.K., Johnston, N., Hutchings, S.W., Gyongy, I., Al~Abbas, T.,
  Dutton, N., Tyler, M., Chan, S., Leach, J.: 256 \texttimes{} 256 40nm/90nm
  {{CMOS 3D-stacked}} 120 {{dB}} dynamic-range reconfigurable time-resolved
  {{SPAD}} imager. In: 2019 {{IEEE International Solid-State Circuits
  Conference-}}({{ISSCC}}). pp. 106--108. {IEEE} (2019).
  \doi{10.1109/ISSCC.2019.8662355}

\bibitem{iams1935secondary}
Iams, H., Salzberg, B.: The secondary emission phototube. Proceedings of the
  Institute of Radio Engineers  \textbf{23}(1),  55--64 (1935)

\bibitem{inglePassiveInterPhotonImaging2021b}
Ingle, A., Seets, T., Buttafava, M., Gupta, S., Tosi, A., Gupta, M., Velten,
  A.: Passive {{Inter-Photon Imaging}}. In: Proceedings of the {{IEEE}}/{{CVF
  Conference}} on {{Computer Vision}} and {{Pattern Recognition}}. pp.
  8585--8595 (2021)

\bibitem{ingleHighFluxPassive2019}
Ingle, A., Velten, A., Gupta, M.: High {{Flux Passive Imaging With
  Single-Photon Sensors}}. In: Proceedings of the {{IEEE}}/{{CVF Conference}}
  on {{Computer Vision}} and {{Pattern Recognition}}. pp. 6760--6769 (2019)

\bibitem{yoshida2021}
{J Yoshida}: {Breaking Down iPad Pro 11's LiDAR Scanner}.
  \url{https://www.eetimes.com/breaking-down-ipad-pro-11s-lidar-scanner/}, {EE
  Times 6/5/2020. Accessed 5/6/2021.}

\bibitem{Jocher_Ultralytics_YOLO_2023}
Jocher, G., Chaurasia, A., Qiu, J.: {Ultralytics YOLOv8} (Jan 2023),
  \url{https://github.com/ultralytics/ultralytics}, {Accessed 3/6/2024}

\bibitem{koppalWideAngleMicrovisionSensors2013}
Koppal, S.J., Gkioulekas, I., Young, T., Park, H., Crozier, K.B., Barrows,
  G.L., Zickler, T.: Toward {{Wide-Angle Microvision Sensors}}. IEEE
  Transactions on Pattern Analysis and Machine Intelligence  \textbf{35}(12),
  2982--2996 (Dec 2013). \doi{10.1109/TPAMI.2013.22}

\bibitem{liuSinglePhotonCameraGuided2022}
Liu, Y., {Gutierrez-Barragan}, F., Ingle, A., Gupta, M., Velten, A.:
  Single-{{Photon Camera Guided Extreme Dynamic Range Imaging}}. In:
  Proceedings of the {{IEEE}}/{{CVF Winter Conference}} on {{Applications}} of
  {{Computer Vision}}. pp. 1575--1585 (2022)

\bibitem{maPhotonnumberresolvingMegapixelImage2017}
Ma, J., Masoodian, S., Starkey, D.A., Fossum, E.R.: Photon-number-resolving
  megapixel image sensor at room temperature without avalanche gain. Optica
  \textbf{4}(12),  1474--1481 (Dec 2017). \doi{10.1364/OPTICA.4.001474}

\bibitem{ma19eRmsRead2021}
Ma, J., Zhang, D., Elgendy, O.A., Masoodian, S.: A 0.19e- rms {{Read Noise}}
  16.{{7Mpixel Stacked Quanta Image Sensor With}} 1.1 {$M$}m-{{Pitch Backside
  Illuminated Pixels}}. IEEE Electron Device Letters  \textbf{42}(6),  891--894
  (Jun 2021). \doi{10.1109/LED.2021.3072842}

\bibitem{maQuantaBurstPhotography2020a}
Ma, S., Gupta, S., Ulku, A.C., Bruschini, C., Charbon, E., Gupta, M.: Quanta
  burst photography. ACM Transactions on Graphics  \textbf{39}(4),  79:1--79:16
  (Aug 2020). \doi{10.1145/3386569.3392470}

\bibitem{maBurstVisionUsing2023}
Ma, S., Mos, P., Charbon, E., Gupta, M.: Burst {{Vision Using Single-Photon
  Cameras}}. In: Proceedings of the {{IEEE}}/{{CVF Winter Conference}} on
  {{Applications}} of {{Computer Vision}}. pp. 5375--5385 (2023)

\bibitem{medinBinomialNegativeBinomial2019}
Medin, S.C., {Murray-Bruce}, J., Casta{\~n}{\'o}n, D., Goyal, V.K.: Beyond
  {{Binomial}} and {{Negative Binomial}}: {{Adaptation}} in {{Bernoulli
  Parameter Estimation}}. IEEE Transactions on Computational Imaging
  \textbf{5}(4),  570--584 (Dec 2019). \doi{10.1109/TCI.2019.2913108}

\bibitem{morimotoMegapixel3DStackedCharge2021}
Morimoto, K., Iwata, J., Shinohara, M., Sekine, H., Abdelghafar, A., Tsuchiya,
  H., Kuroda, Y., Tojima, K., Endo, W., Maehashi, Y., Ota, Y., Sasago, T.,
  Maekawa, S., Hikosaka, S., Kanou, T., Kato, A., Tezuka, T., Yoshizaki, S.,
  Ogawa, T., Uehira, K., Ehara, A., Inui, F., Matsuno, Y., Sakurai, K.,
  Ichikawa, T.: 3.2 {{Megapixel 3D-Stacked Charge Focusing SPAD}} for
  {{Low-Light Imaging}} and {{Depth Sensing}}. In: 2021 {{IEEE International
  Electron Devices Meeting}} ({{IEDM}}). pp. 20.2.1--20.2.4 (Dec 2021).
  \doi{10.1109/IEDM19574.2021.9720605}

\bibitem{morimotoMegapixelTimegatedSPAD2020}
Morimoto, K., Ardelean, A., Wu, M.L., Ulku, A.C., Antolovic, I.M., Bruschini,
  C., Charbon, E.: Megapixel time-gated {{SPAD}} image sensor for {{2D}} and
  {{3D}} imaging applications. Optica  \textbf{7}(4),  346--354 (2020).
  \doi{10.1364/OPTICA.386574}

\bibitem{morimoto2021scaling}
Morimoto, K., Charbon, E.: A scaling law for {SPAD} pixel miniaturization.
  Sensors  \textbf{21}(10), ~3447 (2021)

\bibitem{ogi2023iisw}
Ogi, J., Sano, F., Nakata, T., Kubo, Y., Onishi, W., Koswaththaghe, C.,
  Mochizuki, T., Tashiro, Y., Hizu, K., Takatsuka, T., et~al.: A 3.06 µm spad
  pixel with embedded metal contact and power grid on deep trench pixel
  isolation for high-resolution photon-counting. In: 2023 International Image
  Sensor Workshop ({IISW}) (May 2023)

\bibitem{ogi250fps124dBDynamicRange2021}
Ogi, J., Takatsuka, T., Hizu, K., Inaoka, Y., Zhu, H., Tochigi, Y., Tashiro,
  Y., Sano, F., Murakawa, Y., Nakamura, M., Oike, Y.: A 250fps {{124dB
  Dynamic-Range SPAD Image Sensor Stacked}} with {{Pixel-Parallel Photon
  Counter Employing Sub-Frame Extrapolating Architecture}} for {{Motion
  Artifact Suppression}}. In: 2021 {{IEEE International Solid- State Circuits
  Conference}} ({{ISSCC}}). vol.~64, pp. 113--115 (Feb 2021).
  \doi{10.1109/ISSCC42613.2021.9365977}

\bibitem{ota37W143dBDynamicRange1Mpixel2022}
Ota, Y., Morimoto, K., Sasago, T., Shinohara, M., Kuroda, Y., Endo, W.,
  Maehashi, Y., Maekawa, S., Tsuchiya, H., Abdelahafar, A., Hikosaka, S.,
  Motoyama, M., Tojima, K., Uehira, K., Iwata, J., Inui, F., Matsuno, Y.,
  Sakurai, K., Ichikawa, T.: A 0.{{37W 143dB-Dynamic-Range 1Mpixel
  Backside-Illuminated Charge-Focusing SPAD Image Sensor}} with {{Pixel-Wise
  Exposure Control}} and {{Adaptive Clocked Recharging}}. In: 2022 {{IEEE
  International Solid- State Circuits Conference}} ({{ISSCC}}). vol.~65, pp.
  94--96 (Feb 2022). \doi{10.1109/ISSCC42614.2022.9731644}

\bibitem{po2022adaptive}
Po, R., Pediredla, A., Gkioulekas, I.: Adaptive gating for single-photon 3d
  imaging. In: Proceedings of the {{IEEE}}/{{CVF}} Conference on Computer
  Vision and Pattern Recognition. pp. 16354--16363 (2022)

\bibitem{quSpatiallyVaryingExposure2024}
Qu, X., Chi, Y., Chan, S.H.: Spatially {{Varying Exposure With}} 2-by-2
  {{Multiplexing}}: {{Optimality}} and {{Universality}}. IEEE Transactions on
  Computational Imaging  \textbf{10},  261--276 (2024).
  \doi{10.1109/TCI.2024.3354426}

\bibitem{severini2021spad}
Severini, F., Cusini, I., Berretta, D., Pasquinelli, K., Incoronato, A., Villa,
  F.: {SPAD} pixel with sub-ns dead-time for high-count rate applications. IEEE
  Journal of Selected Topics in Quantum Electronics  \textbf{28}(2: Optical
  Detectors), ~1--8 (2021)

\bibitem{shenzhencmtechnologycompanyltd16MPMIPI2023}
{Shenzhen CM Technology company Ltd}: 16 {{MP MIPI Camera Module}} with {{SONY
  IMX206}} sensor.
  http://www.camera-module.com/product/mipicameramodule/16mp-mipi-camera-module-sony-imx206-sensor.html
  (Nov 2023)

\bibitem{Sundar:2023:SoDaCam}
Sundar, V., Ardelean, A., Swedish, T., Brusschini, C., Charbon, E., Gupta, M.:
  Sodacam: Software-defined cameras via single-photon imaging. In: Proceedings
  of the {IEEE} {International} {Conference} on {Computer} {Vision} ({ICCV})
  (2023)

\bibitem{takatsuka36UmpitchSPAD2023}
Takatsuka, T., Ogi, J., Ikeda, Y., Hizu, K., Inaoka, Y., Sakama, S., Watanabe,
  I., Ishikawa, T., Shimada, S., Suzuki, J., Maeda, H., Toshima, K., Nonaka,
  Y., Yamamura, A., Ozawa, H., Koga, F., Oike, Y.: A 3.36 {$\mu$}m-pitch
  {{SPAD}} photon-counting image sensor using clustered multi-cycle clocked
  recharging technique with intermediate most-significant-bit readout. In: 2023
  {{IEEE Symposium}} on {{VLSI Technology}} and {{Circuits}} ({{VLSI
  Technology}} and {{Circuits}}). pp.~1--2 (Jun 2023).
  \doi{10.23919/VLSITechnologyandCir57934.2023.10185241}

\bibitem{takatsuka36MuMpitch2023}
Takatsuka, T., Ogi, J., Ikeda, Y., Hizu, K., Inaoka, Y., Sakama, S., Watanabe,
  I., Ishikawa, T., Shimada, S., Suzuki, J., Maeda, H., Toshima, K., Nonaka,
  Y., Yamamura, A., Ozawa, H., Koga, F., Oike, Y.: A 3.36
  \${\textbackslash}mu\$m-pitch {{SPAD}} photon-counting image sensor using
  clustered multi-cycle clocked recharging technique with intermediate
  most-significant-bit readout. In: 2023 {{IEEE Symposium}} on {{VLSI
  Technology}} and {{Circuits}} ({{VLSI Technology}} and {{Circuits}}).
  pp.~1--2 (Jun 2023). \doi{10.23919/VLSITechnologyandCir57934.2023.10185241}

\bibitem{tilmonEnergyEfficientAdaptive3D2023a}
Tilmon, B., Sun, Z., Koppal, S.J., Wu, Y., Evangelidis, G., Zahreddine, R.,
  Krishnan, G., Ma, S., Wang, J.: Energy-{{Efficient Adaptive 3D Sensing}}. In:
  Proceedings of the {{IEEE}}/{{CVF Conference}} on {{Computer Vision}} and
  {{Pattern Recognition}}. pp. 5054--5063 (2023)

\bibitem{ulku512512SPAD2019}
Ulku, A.C., Bruschini, C., Antolovic, I.M., Kuo, Y., Ankri, R., Weiss, S.,
  Michalet, X., Charbon, E.: A 512 \texttimes{} 512 {{SPAD Image Sensor With
  Integrated Gating}} for {{Widefield FLIM}}. IEEE Journal of Selected Topics
  in Quantum Electronics  \textbf{25}(1),  1--12 (Jan 2019).
  \doi{10.1109/jstqe.2018.2867439}

\bibitem{wang2007laplacian}
Wang, X.: Laplacian operator-based edge detectors. IEEE Transactions on Pattern
  Analysis and Machine Intelligence  \textbf{29}(5),  886--890 (2007)

\bibitem{wang2004image}
Wang, Z., Bovik, A.C., Sheikh, H.R., Simoncelli, E.P.: Image quality
  assessment: From error visibility to structural similarity. IEEE Transactions
  on Image Processing  \textbf{13}(4),  600--612 (2004)

\bibitem{wayne2021low}
Wayne, M.A., Bienfang, J.C., Migdall, A.L.: Low-noise photon counting above
  100$\times$106 counts per second with a high-efficiency reach-through
  single-photon avalanche diode system. Applied Physics Letters
  \textbf{118}(13) (2021)

\bibitem{xie15hed}
Xie, S., Tu, Z.: Holistically-nested edge detection. In: Proceedings of
  {{IEEE}} International Conference on Computer Vision. pp. 1395--1403 (2015)

\bibitem{xu2020compact}
Xu, Y., Lu, J., Wu, Z.: A compact high-speed active quenching and recharging
  circuit for {{SPAD}} detectors. IEEE Photonics Journal  \textbf{12}(5), ~1--8
  (2020)

\bibitem{yang2011bits}
Yang, F., Lu, Y.M., Sbaiz, L., Vetterli, M.: Bits from photons: Oversampled
  image acquisition using binary poisson statistics. IEEE Transactions on image
  processing  \textbf{21}(4),  1421--1436 (2011)

\end{thebibliography}
}

\end{document}